\documentclass[letterpaper,titlepage,11pt]{article}
\usepackage[colorlinks=true,urlcolor=blue,citecolor=magenta]{hyperref}
\usepackage{amssymb,amsmath,amsfonts}
\usepackage{epsfig}
\usepackage{epsfig}
\usepackage{graphicx}
\usepackage{epstopdf}
\def\clock{{\count0=\time
           \divide\count0 60
           \ifnum\count0<10 0\fi\the\count0
           \multiply\count0 -60 \advance\count0 \time
           :\ifnum\count0<10 0\fi \the\count0
         }}
\newcommand{\timestamp}{{\small\vbox{\hbox{\tt\jobname.tex}
\hbox{\the\day/\the\month/\the\year, \clock}}}}

\usepackage{amsmath}
\usepackage{amssymb}
\usepackage{amsfonts}
\usepackage{amscd}
\usepackage{amsthm}
\usepackage{latexsym}
\usepackage{amsbsy}
\usepackage{bbm}
\usepackage[english]{babel}
\usepackage{graphicx}
\usepackage{psfrag}

\newcommand{\spa}{\quad , \quad} 

\numberwithin{equation}{section}
\begin{document}

\begin{titlepage}

\phantom{hello}
 \vskip 2.0 cm

\centerline{\Huge \bf 
Holographic Models for Theories with } \vskip .2cm  \centerline{\Huge \bf Hyperscaling Violation
}

\vskip 1.5cm

\centerline{\large {{\bf Jakob Gath$^a$, Jelle Hartong$^a$, Ricardo Monteiro$^b$, Niels A. Obers$^a$}}} 
\vspace{1cm}

\begin{center}
 $^a${\small\slshape Niels Bohr Institute,\\
 University of Copenhagen,\\
 Blegdamsvej 17, DK-2100 Copenhagen \O,
 Denmark}
 \vskip .2cm
 $^b${\small\slshape Niels Bohr International Academy and Discovery Center,\\
Niels Bohr Institute,\\ 
University of Copenhagen,\\
Blegdamsvej 17, DK-2100 Copenhagen \O, Denmark} 
 
 \end{center}
 
 \vspace{0.2cm}

\centerline{{\small \texttt{gath@nbi.dk, hartong@nbi.dk, monteiro@nbi.dk, obers@nbi.dk}}}

\vskip 1.3cm \centerline{\bf Abstract} \vskip 0.2cm \noindent

We study in detail a variety of gravitational toy models for hyperscaling-violating Lifshitz (hvLif) space-times. These
space-times have been recently explored as holographic dual models for
condensed matter systems. We start by considering a model of
gravity coupled to a massive vector field and a dilaton with a potential.
This model supports the full class of hvLif space-times and special attention is
given to the particular values of the scaling exponents
appearing in certain non-Fermi liquids. We study linearized perturbations
in this model, and consider probe fields whose interactions mimic those of
the perturbations. The resulting equations of motion for the probe fields are invariant under the Lifshitz scaling. 
We derive Breitenlohner--Freedman-type bounds for these new probe fields. For the cases of interest 
the hvLif space-times have curvature invariants that blow up in the UV. We study the problem of constructing models in 
which the hvLif space-time can have an AdS or Lifshitz UV completion. We also analyze reductions of Schr\"odinger space-times and
reductions of waves on extremal (intersecting) branes, accompanied by transverse space reductions, that are solutions to supergravity-like theories,
exploring the allowed parameter range of the hvLif scaling exponents.

\end{titlepage}

\newpage

\begingroup
\hypersetup{linkcolor=black}
\tableofcontents
\endgroup

%%%%%%%%%%%%%%%%%%%%%%%%%%%%%%%%%%
%%%%%%%%%%%%%%%%%%%%%%%%%%%%%%%%%%
\section{Introduction}

Applications of the holographic correspondence to real-world field theories have led to a renewed interest in the phenomenology of gravitational models. According to the duality, the phases of a strongly-coupled quantum field theory may be described by gravity at weak coupling, supplemented by appropriate matter fields. The most established examples of holography involve conformal field theories in $d+1$ dimensions dual to string theory (at low energies, Einstein gravity and matter) in anti-de-Sitter space-times with $d+2$ dimensions. However, there is a wide-ranging effort to understand and make use of holography in a more generic way. The application of holographic techniques to the study of condensed matter systems (see 
Refs.~\cite{Hartnoll:2009sz,Herzog:2009xv,McGreevy:2009xe,Sachdev:2011wg} for reviews) 
 is one of the most active fronts in this development.
 
The goal of this paper is to provide a tentative survey of gravitational toy models which possess certain properties of interest to the study of condensed matter systems. In particular, we will consider toy models which admit metrics of the form
\begin{equation}
\label{dsd}
ds^2=r^{2\theta/d}\left(-\frac{dt^2}{r^{2z}}+\frac{dr^2}{r^2}+\frac{dx_i^2}{r^2}\right)\,.
\end{equation}
which we refer to  as hyperscaling-violating Lifshitz (hvLif) space-times. 
This geometry has two independent parameters, the ``dynamical critical exponent" $z$ and the ``hyperscaling violation exponent" $\theta$, and possesses the scaling property
\begin{equation}
\label{Lscaling}
r \to \lambda r \spa  x_i \to \lambda\, x_i \spa 
t \to \lambda^z\, t \spa ds^2  \to \lambda^{2 \theta/d}\, ds^2 \,  . 
\end{equation}
When $\theta=0$, $z=1$, the solution is simply anti-de-Sitter space-time in Poincar\'e coordinates, the most familiar setting for holography. When $\theta=0$ but $z\neq 1$, we have a so-called Lifshitz (Lif) space-time, since the solution has a non-relativistic scaling symmetry of Lifshitz type, parametrized by $z$. This type of scaling is natural at critical points in several condensed matter models. Therefore, Lifshitz space-times have been proposed as gravity duals of these critical points \cite{Kachru:2008yh,Taylor:2008tg,Goldstein:2009cv}, and the problem of embedding these space-times into string theory has been 
studied in several works \cite{Azeyanagi:2009pr,Hartnoll:2009ns,Balasubramanian:2010uk,Donos:2010tu,Gregory:2010gx,Donos:2010ax,Cassani:2011sv,Halmagyi:2011xh,Narayan:2011az,Chemissany:2011mb}.

Our analysis is based on three perspectives on hvLif space-times: 
\begin{itemize}
\item Lifshitz perspective
\item Schr\"odinger perspective
\item waves-on-branes  perspective
\end{itemize}
with emphasis on the first. The first takes as a starting point (further motivated below)  a model that we refer to as Einstein-Proca-dilaton (EPD) which is gravity coupled to a massive vector field and a scalar field with a potential. 
 We find that this is a natural toy model for the full class of hvLif space-times and
in particular we identify four-dimensional solutions that allow for the exponents $\theta = 1$, $z=3/2$. 
These involve negative mass squared
for the vector field and by performing a linearized perturbation analysis we argue that these do not lead to instabilities. Furthermore we use this analysis to construct a new class of probe fields in hvLif space-times, 
whose couplings mimic those of the linearized perturbations and whose equations of motion
are invariant under the Lifshitz scaling \eqref{Lscaling}. This allows us  to discuss Breitenlohner-Freedman (BF)-like bounds and as a potential application to consider their two-point functions in hvLif space-times. 

The Lifshitz perspective is further explored by examining a more general class of models that admit both hvLif space-times as well as AdS or Lif space-times, which allows for the existence of interpolating solutions with an AdS/Lif in the ultraviolet (UV) and hvLif in the infrared (IR). In particular, this class of models exhibits the novel feature that a quartic
interaction term for the massive vector field is introduced. 

We briefly consider two other perspectives.  The Schr\"odinger perspective starts with 
the observation that the natural theory for this class is gravity coupled to a massive vector field, and by dimensional
reduction these give rise to hvLif space-times.  
Finally, we examine a waves-on-branes perspective which is in part motivated by the 
Schr\"odinger perspective,  and at the same time inspired by the fact that these models naturally occur
in a supergravity/string theory setting. The corresponding actions arise from reducing higher-dimensional 
gravity coupled to dilatons and higher rank gauge potentials. 

The motivation for studying hvLif spaces comes from the precise way in which hyperscaling is violated
and the behavior of the entanglement entropy.  
Hyperscaling is the property that the thermal entropy scales with the
temperature to the power $d/z$ where $d$ is the number of spatial dimensions
and $z$ the dynamical critical exponent. Black holes in AdS ($z=1$) or Lifshitz ($z>1$)
space-times satisfy this property. The entanglement entropy for such
space-times scales with the area of the boundary entangling region.
For hvLif space-times one finds a  scaling for the entropy density with respect to the temperature of the type $S\sim T^{(d-\theta)/z}$ where $\theta$ describes the violation of hyperscaling \cite{Fisher:1986zz}{}\footnote{Hyperscaling occurs when the scaling is determined by the dynamical critical exponent $z$, that is, $\theta=0$.}. The exponent $\theta$ parametrizes the scaling of the proper distance, which is holographically connected to the thermal entropy density in the dual theory. There is a physical condition that, for any local quantum field theory, the entanglement entropy of a certain region scales with the area of the boundary of that region. In terms of the exponent $\theta$, this condition translates into
\begin{equation}
\theta \leq d-1\,.
\end{equation}
When this inequality is saturated, the area law is only valid up to a logarithmic correction, which is consistent with the presence of a co-dimension one Fermi surface \cite{Ogawa:2011bz,Huijse:2011ef}. This ties in well with the fact that for $\theta=d-1$ the thermal entropy scales like $S\sim T^{1/z}$ and hence `sees' an effective one-dimensional system (the direction transverse to the Fermi surface) of thermal excitations with dynamical exponent $z$.

There is also a constraint for the dynamical exponent $z$ in the holographic set-up. This constraint is the null energy condition, which any sensible gravitational system must satisfy. It implies that
\begin{equation}
z \geq 1 + \frac{\theta}{d}\,.
\end{equation}
The simultaneous saturation of the two inequalities leads for a given $d$ to fixed values of $\theta$ and $z$ of particular interest. Remarkably,
these values are precisely the ones relevant for gauge theories of non-Fermi liquid states in $d = 2$ \cite{Ogawa:2011bz}, giving the exponents
\begin{equation}
\label{thetaz}
\theta=1\,, \qquad z =\frac{3}{2}\,.
\end{equation}
There has been considerable interest in the analysis of toy models and string theory embeddings admitting metrics of the type \eqref{dsd}. This work started with \cite{Charmousis:2010zz,Iizuka:2011hg,Gouteraux:2011ce,Huijse:2011ef,Ogawa:2011bz,Dong:2012se}, and proceeded in 
\cite{Singh:2010zs,Narayan:2012hk,Singh:2012un,Dey:2012tg,Dey:2012rs,Charmousis:2012dw,
Ammon:2012je,Bhattacharya:2012zu,Kundu:2012jn,Dey:2012fi,Bueno:2012sd}{$\,$}\footnote{See also 
Refs.~ \cite{Shaghoulian:2011aa,Hartnoll:2012wm,Takayanagi:2012kg,Dey:2012hf,Kulaxizi:2012gy,Alishahiha:2012cm,Edalati:2012tc} for more
on entanglement entropy and other physical properties of hvLif space-times.}.  
The present paper puts some of these approaches in a more general setting. Moreover, explicit well-defined toy models admitting the special exponents \eqref{thetaz} will be proposed here for the first time, to the best of our knowledge.

Before we proceed, it is important to comment on the infrared (IR, $r\to \infty$) and ultraviolet (UV, $r\to 0$) limits of the space-times \eqref{dsd}.\footnote{We consider here that $\theta \leq d$, such that the hyperscaling violation $S\sim T^{(d-\theta)/z}$ does not lead to a violation of the third law of thermodynamics.}
Lifshitz space-times are known to be singular in the IR, where they possess a null curvature singularity where tidal forces diverge \cite{Kachru:2008yh,Copsey:2010ya,Horowitz:2011gh}. The introduction of hyperscaling violation complicates the situation generically, as the infrared may possess a naked singularity with diverging curvature invariants for $\theta <0$ and/or divergent tidal forces
\cite{Shaghoulian:2011aa,Copsey:2012gw}. The exception occurs precisely, amongst others, for the special exponents \eqref{thetaz}, in which case the IR limit of the metric is  regular. As for the UV limit, the cases with $\theta > 0$ have singular curvature invariants. Generically, a metric of the type \eqref{dsd} should be taken as an effective metric in the ``interior" of the space-time, away from the infrared and ultraviolet limits, as emphasised in \cite{Dong:2012se}. The problem of regularization of the IR and UV regions in Lifshitz and hvLif holography has been discussed in several scenarios in the literature \cite{Adams:2008zk,Gouteraux:2011ce,Harrison:2012vy,Charmousis:2012dw,Bhattacharya:2012zu,Kundu:2012jn}.

An outline of the paper and summary of the main results is as follows: 

We start in Section 2 by introducing the EPD model consisting of gravity coupled to a vector field with
a mass  term, along with a scalar field with a potential term, which in our concrete applications  is of exponential form. This choice of model is motivated by the fact that Lif space-times occur in two generic types of toy
models: One including a  massive vector field, the other including a massless vector and a dilaton. 
Only in the former case does the entire background preserve the Lifshitz symmetries, while in the latter one
they are broken by the matter fields.   The EPD model with a mass term $m$ for the vector is thus a natural
starting point\footnote{Even though the dual field theory interpretation of the fields in our model depend on the explicit UV completion we expect that the case of a positive mass for the vector field corresponds to an explicit or spontaneous breaking of a $U(1)$ symmetry.} for our analysis, and for concreteness we focus on the case $d=2$. 
In particular, turning on a mass term we find that this model admits the entire class of hvLif space-times, provided one
allows $m^2$ to be negative, but bounded from below.  In these solutions the scalar runs logarithmically with the radial distance $r$.  We also identify the conditions
such that scale invariance is restored when $\theta \rightarrow 0$, i.e. the breaking of scale
invariance both for the metric and all matter fields is proportional to $\theta$.  

The class of models with $m^2 \geq 0$ fails to describe  the two NEC respecting cases $\theta=2$ ($z >0$ or $z <1$) and $\theta = 2(z-1)$ ($ 1 < z < 4$), which includes in particular the phenomenologically interesting case \eqref{thetaz}. We first identify the general properties that the energy momentum tensor of the matter sector
should satisfy in order to support such solutions. In particular, we find that for $\theta = 2(z-1)$ the weak
energy condition (WEC) is violated, due to the fact that $m^2 < 0$ (and the potential being negative).

The fact that we find that  $m^2 <0$ in the above-mentioned cases, raises the question of whether
the theory has an instability. To this end we 
perform in Section 3 a linearized perturbation analysis around our hvLif backgrounds, focussing on the purely
radial perturbations which are the lowest lying modes of the system. As a result we find that
for a very large range of $\theta$, $z$ values, including the special case \eqref{thetaz}, there is no instability. 
As an application of this analysis, we construct new scalar and vector probe fields whose couplings to the background mimic those of the perturbations and for which we derive BF-like bounds.

A generalization of the EPD model that allows for a non-negative mass squared for the vector field describing also the case \eqref{thetaz} is discussed in Subsections \ref{subsec:ViolatingWEC} and \ref{subsec:solutions} and consists of an action with an additional quartic couping for the vector field. Such terms appear naturally in the context of Lorentz violating field theories \cite{Kiritsis:2012ta}.

As mentioned before hvLif spaces of interest (e.g. those with $\theta=1$) are UV singular, so one might want to  
replace the UV with an  AdS or Lifshitz space-time. This would require a solution that interpolates from an
IR hvLif to a UV AdS or Lif space-time. For explicitness we assume that the action describing the UV solution is the same as the one describing the IR solution. In more realistic scenarios this need not be the case. We then proceed to require that both the UV and the IR are themselves solutions of the equations of motion. Such a setup guarantees that the UV and IR geometries decouple from the full interpolating solutions so that one can perturb around these geometries in a well-defined way. We expect that in this case there exists a well-defined limit for a probe field on the full interpolating solution that reduces to the solution for a probe field on the UV or IR geometries. We therefore classify in Section 4 a particular set of Lagrangians that have a $\theta=1$ hvLif in the IR and AdS or Lif in the UV, both as solutions to the equations of motion. 
This is achieved in the above-mentioned generalization of the EPD model with quartic couplings for the vector field. In particular, we consider a multi-scalar model with a constant potential and show that one can have $\theta=1$ in the IR and $\theta=0$ in the UV for large ranges of their respective $z$ values. 

We then briefly discuss in Section 5 a slightly different approach that is also based on massive vector fields, but
which uses dimensional reduction of a Schr\"odinger space-time.  This leads to a class of models with hvLif
solutions where we have both dilatonic and axionic scalars.  Motivated by the fact that Schr\"odinger space-times can be viewed as a wave-type deformation of AdS, we consider in Section 6 waves propagating on extremal branes, where the AdS factor arises in the near-horizon limit.  The starting point here is gravity coupled to dilatons and higher rank gauge potentials in dimension $D > d+2$ and the corresponding extremal brane solutions,  including
intersecting branes. By adding a wave propagating along the world volume and subsequently reducing 
along the transverse space of the brane as well as the direction of the wave, we then construct hvLif spaces
as solutions of the corresponding reduced actions. We expect this class of models to be relevant in finding
hvLif solutions in string theory, and more generally supergravity. 
\vskip .5cm
\subsubsection*{Note added}

While this work was being completed the papers 
\cite{Iizuka:2012pn,Gouteraux:2012yr} appeared on the arXiv. They contain some overlap with this
manuscript.

%%%%%%%%%%%%%%%%%%%%%%%%%%%%%%%%%%
%%%%%%%%%%%%%%%%%%%%%%%%%%%%%%%%%%
\section{A Lifshitz perspective}

Lifshitz space-times are solutions of the type \eqref{dsd} with $\theta=0$, which must be supported by matter fields. Two generic types of toy models have been proposed for these space-times: one including a massive vector field; the other including a massless vector and a dilaton. In the latter case, the matter fields necessarily break the Lifshitz scale invariance, as mentioned in the introduction. We therefore start by considering hvLif solutions of the EPD model. Our main result is the solution 
\eqref{eq:phim^2neq0}--\eqref{eq:am^2neq0}.

\subsection{EPD model and equations of motion}

Consider the following four-dimensional theory
\begin{equation}\label{eq:Lifmodel}
S=\int d^4x\sqrt{-g}\left(R-\frac{1}{2}(\partial\phi)^2-\frac{1}{4}e^{a\phi}F^2-\frac{m^2}{2}e^{b\phi}A^2-V(\phi)\right)\,.
\end{equation}
For $m^2=0$ this is known as the Einstein--Maxwell--dilaton (EMD) model, and for $m^2 \neq 0$ we will refer to this as the Einstein--Proca--dilaton (EPD) model.
 The equations of motion are
\begin{eqnarray}
G_{\mu\nu} & = & \frac{1}{2}\left[\partial_\mu\phi\partial_\nu\phi-\frac{1}{2}(\partial\phi)^2g_{\mu\nu}+e^{a\phi}\left(F_{\mu\rho}F_\nu{}^\rho-\frac{1}{4}F^2g_{\mu\nu}\right)\right.\nonumber\\
&&\left.+m^2e^{b\phi}\left(A_\mu A_\nu-\frac{1}{2}A^2g_{\mu\nu}\right)-Vg_{\mu\nu}\right]\,,\label{eq:Einsteineqs}\\
\square\phi & = & \frac{a}{4}e^{a\phi}F^2+\frac{b}{2}m^2e^{b\phi}A^2+\frac{dV}{d\phi}\,,\\
\nabla_\mu (e^{a\phi}F^{\mu\nu}) & = & m^2e^{b\phi}A^\nu\,.
\end{eqnarray}
We want to know when this model admits a metric solution of the type
\begin{equation}
\label{dsd4}
ds^2=r^\theta\left(-\frac{dt^2}{r^{2z}}+\frac{dr^2}{r^2}+\frac{dx^2+dy^2}{r^2}\right)\,.
\end{equation}
The nonzero components of the Einstein equations are
\begin{eqnarray}
G_{tt} & = & -\frac{1}{4}r^{-2z}\left(\theta^2-8\theta+12\right)\,,\\
G_{rr} & = & \frac{1}{4}r^{-2}\left(3\theta^2-4z\theta-8\theta+8z+4\right)\,,\\
G_{xx} & = & G_{yy}=\frac{1}{4}r^{-2}\left(\theta^2-4z\theta-4\theta+4z^2+4z+4\right)\,.
\end{eqnarray}
The null energy condition $G_{\mu\nu}\xi^\mu\xi^\nu\ge 0$ for every $\xi^\mu$ with $g_{\mu\nu}\xi^\mu\xi^\nu=0$ gives
\begin{eqnarray}
0 & \le & (z-1)(z+2-\theta)\,,\label{eq:NEC1}\\
0 & \le & (\theta-2)(\theta+2-2z)\,.\label{eq:NEC2}
\end{eqnarray}
We assume that the scalar field $\phi$ is only a function of $r$. For $m^2\neq 0$, we require
\begin{equation}
A=A_t(r)dt\,.
\end{equation}
This comes from imposing the symmetries $\mathcal{L}_KA_{\mu}=0$, with $K\in\{\partial_t,\partial_x,\partial_y,x\partial_y-y\partial_x\}$ as well as the
$r$-component of the $A_\mu$ equation of motion. For $m^2=0$, we take instead
\begin{equation}
F=F_{rt}(r)dr\wedge dt+Pdx\wedge dy\,,
\end{equation}
where $P$ is a constant. This comes from imposing the symmetries $\mathcal{L}_KF_{\mu\nu}=0$, with $K\in\{\partial_t,\partial_x,\partial_y,x\partial_y-y\partial_x\}$, and also $dF=0$.

When $m^2\neq 0$, the equations of motion with the above Ans\"atze become
\begin{eqnarray}
r^2\phi'^2+r^{2+2z-\theta}e^{a\phi}A_t'^2+m^2r^{2z}e^{b\phi}A_t^2+2r^\theta V & = & -\theta^2+8\theta-12\,,\label{eq:Einsteqtt}\\
r^2\phi'^2-r^{2+2z-\theta}e^{a\phi}A_t'^2+m^2r^{2z}e^{b\phi}A_t^2-2r^\theta V & = & 3\theta^2-4z\theta-8\theta+8z+4\,,\label{eq:Einsteqrr}\\
-r^2\phi'^2+r^{2+2z-\theta}e^{a\phi}A_t'^2+m^2r^{2z}e^{b\phi}A_t^2-2r^\theta V & = & \theta^2-4z\theta-4\theta+4z^2+4z+4\,,\label{eq:Einsteqxx}
\end{eqnarray}
\begin{equation}\label{eq:eomphi}
r^{-\theta+z+3}\frac{d}{dr}\left(r^{\theta-z-1}\phi'\right)=-\frac{a}{2}r^{2+2z-\theta}e^{a\phi}A_t'^2-\frac{b}{2}m^2r^{2z}e^{b\phi}A_t^2+r^\theta\frac{dV}{d\phi}\,,
\end{equation}
\begin{equation}\label{eq:eomA}
r^{-z+3}\frac{d}{dr}\left(r^{z-1}e^{a\phi}A_t'\right)=m^2r^\theta e^{b\phi}A_t\,.
\end{equation}
Adding \eqref{eq:Einsteqtt} and \eqref{eq:Einsteqrr} gives
\begin{equation}
r^2\phi'^2+m^2r^{2z}e^{b\phi}A_t^2=(\theta-2)(\theta+2-2z)\,,
\end{equation}
while adding \eqref{eq:Einsteqtt} and \eqref{eq:Einsteqxx} yields
\begin{equation}
r^{2+2z-\theta}e^{a\phi}A_t'^2+m^2r^{2z}e^{b\phi}A_t^2=2(z-1)(z+2-\theta)\,.
\end{equation}
Note that the NEC conditions \eqref{eq:NEC1} and \eqref{eq:NEC2} appear on the right hand side of these equations. Adding \eqref{eq:Einsteqrr} and \eqref{eq:Einsteqxx} gives
\begin{equation}
m^2r^{2z}e^{b\phi}A_t^2-2r^\theta V=2(z+1-\theta)(z+2-\theta)\,.
\end{equation}
In this subsection we only consider the case $m^2>0$. In the next subsection we will consider negative $m^2$.

When $m^2=0$, the equations of motion with the above Ans\"atze become
\begin{eqnarray}
r^2\phi'^2+r^{2+2z-\theta}e^{a\phi}A_t'^2+P^2r^{4-\theta}e^{a\phi}+2r^\theta V & = & -\theta^2+8\theta-12\,,\label{eq:Einsteqttm2=0}\\
r^2\phi'^2-r^{2+2z-\theta}e^{a\phi}A_t'^2-P^2r^{4-\theta}e^{a\phi}-2r^\theta V & = & 3\theta^2-4z\theta-8\theta+8z+4\,,\label{eq:Einsteqrrm2=0}\\
-r^2\phi'^2+r^{2+2z-\theta}e^{a\phi}A_t'^2+P^2r^{4-\theta}e^{a\phi}-2r^\theta V & = & \theta^2-4z\theta-4\theta+4z^2+4z+4\,,\label{eq:Einsteqxxm2=0}
\end{eqnarray}
\begin{equation}
r^{-\theta+z+3}\frac{d}{dr}\left(r^{\theta-z-1}\phi'\right)=-\frac{a}{2}r^{2+2z-\theta}e^{a\phi}A_t'^2+\frac{a}{2}P^2r^{4-\theta}e^{a\phi}+r^\theta\frac{dV}{d\phi}\,,
\end{equation}
\begin{equation}
\frac{d}{dr}\left(r^{z-1}e^{a\phi}A_t'\right)=0\,.
\end{equation}
Adding \eqref{eq:Einsteqttm2=0} and \eqref{eq:Einsteqrrm2=0} gives
\begin{equation}
r^2\phi'^2=(\theta-2)(\theta+2-2z)\,,
\end{equation}
while adding \eqref{eq:Einsteqttm2=0} and \eqref{eq:Einsteqxxm2=0} yields
\begin{equation}
r^{2+2z-\theta}e^{a\phi}A_t'^2+P^2r^{4-\theta}e^{a\phi}=2(z-1)(z+2-\theta)\,.
\end{equation}
Note that the NEC conditions \eqref{eq:NEC1} and \eqref{eq:NEC2} appear on the right hand side of these equations. Adding \eqref{eq:Einsteqrrm2=0} and \eqref{eq:Einsteqxxm2=0} gives
\begin{equation}
V=-(z+1-\theta)(z+2-\theta)r^{-\theta}\,.
\end{equation}

\subsection{Solutions}

We will now consider some particular cases of solutions to these equations of motion both for $m^2=0$ and $m^2\neq 0$. The novel solution, which will be the focus of our attention in this paper, is presented in Eqs.~\eqref{eq:phim^2neq0}--\eqref{eq:am^2neq0}.

\subsubsection*{Zero vector field}

We will first demand that $A_\mu=0$. This is equivalent to setting $z=1$ or $z=\theta-2$, so that one of the two conditions in \eqref{eq:NEC1} and \eqref{eq:NEC2} is saturated. 
Let us start with setting $z=1$. This requires (regardless the value of $m^2$)
\begin{eqnarray}
\phi & = & \phi_0+\alpha\log r\,,\\
V & = & -(\theta-2)(\theta-3)e^{-\frac{\theta}{\alpha}(\phi-\phi_0)}\,,\\
\alpha^2 & = & \theta(\theta-2)\,,\\
A_\mu & = & 0\,.
\end{eqnarray}
On the other hand, setting $z=\theta-2$ requires (regardless the value of $m^2$)
\begin{eqnarray}
\phi & = & \phi_0+\alpha\log r\,,\\
V & = & 0\,,\\
\alpha^2 & = & -(\theta-6)(\theta-2)\,,\\
A_\mu & = & 0\,.
\end{eqnarray}
 The category with $z=1$ of course contains the AdS solution $(\theta,z)=(0,1)$ whereas the category $z=\theta-2$ contains the Ricci-flat case $(\theta,z)=(6,4)$ \cite{Bueno:2012sd}; the latter category is to be disregarded, however, as it can be shown not to describe an extremal horizon.

\subsubsection*{Constant scalar field} 

In the $m^2=0$ case there is only one case not already contained in the zero vector case dicussed above that has a constant scalar field. This solution has $(\theta,z)=(4,3)$ and is given by
\begin{eqnarray}
\phi & = & \phi_0\,,\\
V & = & 0\,,\\
F_{rt} & = & Qe^{-a\phi_0}r^{-2}\,,
\end{eqnarray}
with either $a=0$ and $Q^2+P^2=4$ or $Q^2e^{-a\phi_0}=P^2e^{a\phi_0}=2$ (see also \cite{Bueno:2012sd}). It should be disregarded as it can be shown not to describe an extremal horizon.

In the $m^2\neq 0$ case, putting the scalar equal to a constant leads to the following class of $\theta=0$ (Lifshitz and AdS) solutions ($z\ge 1$)
\begin{eqnarray}
\phi & = & \phi_0\,,\label{eq:Lifshitz1}\\
V & = & -(z^2+z+4)\,,\\
A_t & = & \pm e^{-a\phi_0/2}\left(\frac{2(z-1)}{z}\right)^{1/2}r^{-z}\,,\\
m^2 & = & 2ze^{(a-b)\phi_0}\,,\\
b & = & -az/2\,.\label{eq:Lifshitz5}
\end{eqnarray}

\subsubsection*{Massless vector field}  

We first consider the $m^2=0$ case. We must have
\begin{eqnarray}
\phi & = & \phi_0+\alpha\log r\,,\label{eq:phim=0case}\\
V & = & -(z+1-\theta)(z+2-\theta)e^{-\tfrac{\theta}{\alpha}(\phi-\phi_0)}\,,\\
\alpha^2 & = & (\theta-2)(\theta+2-2z)>0\,.\label{eq:xm=0case}
\end{eqnarray}
There are three classes of solutions. The first class is 
\begin{eqnarray}
a & = & \frac{1}{\alpha}(4-\theta)\neq 0\,,\\
P & = & 0\,,\\
Q^2 & = & 2e^{\tfrac{1}{\alpha}(4-\theta)\phi_0}(z-1)(z+2-\theta)>0\,,\\
F_{rt} & = & Qe^{-\tfrac{1}{\alpha}(4-\theta)\phi_0}r^{\theta-z-3}\,,\label{eq:electricflux}
\end{eqnarray}
the second class has 
\begin{eqnarray}
a & = & -\frac{1}{\alpha}(4-\theta)\neq 0\,,\\
P^2 & = & 2e^{\tfrac{1}{\alpha}(4-\theta)\phi_0}(z-1)(z+2-\theta)>0\,,\\
Q & = & 0\,,\\
F_{rt} & = & 0\,,
\end{eqnarray}
and finally for the third class 
\begin{eqnarray}
a & = & 0\,,\\
\theta & = & 4\,,\\
P^2+Q^2 & = & 2(z-1)(z-2)>0\,,\\
F_{rt} & = & Qr^{-z+1}\,.
\end{eqnarray}
For the third class, we also have $\alpha^2=4(3-z)>0$. The first and second classes are related by electric/magnetic (EM) duality, whereas the third class is EM self-dual. Furthermore, for the first class we have under the Lifshitz scaling
\begin{equation}\label{eq:scaling}
r\rightarrow\lambda r\,,\qquad t\rightarrow\lambda^z t\,,\qquad\vec x\rightarrow\lambda\vec x\,,
\end{equation}
for the 2-form field strength $F\rightarrow\lambda^{\theta-2}F$, while for the second and third classes we have the scaling $F\rightarrow\lambda^2 F$. Note that it is possible to have $\theta=0$ in which case we have for the first class 
\begin{eqnarray}
\phi & = & \phi_0+\alpha\log r\,,\label{eq:Lifm^2=0_1}\\
V & = & -(z+1)(z+2)\,,\\
\alpha^2 & = & 4(z-1)\,,\\
a & = & \frac{4}{\alpha}\,,\\
P & = & 0\,,\\
Q^2 & = & 2e^{\tfrac{4}{\alpha}\phi_0}(z-1)(z+2)\,,\\
F_{rt} & = & Qe^{-\tfrac{4}{\alpha}\phi_0}r^{-z-3}\,.\label{eq:Lifm^2=0_7}
\end{eqnarray}
This is the Lifshitz solution of the EMD model where the matter fields break the Lifshitz scale invariance as opposed to the Lifshitz solution \eqref{eq:Lifshitz1}--\eqref{eq:Lifshitz5} of the EPD model.

\subsubsection*{Massive vector field}  

We now come to our main result regarding hvLif solutions. For this we consider the case $m^2\neq 0$ and assume the simplest form of the scalar field allowing for non-trivial solutions
\begin{equation}\label{eq:phim^2neq0}
\phi = \phi_0+\alpha \log r\,.
\end{equation}
In the $m^2=0$ case this was the only possibility. It is not clear at this stage if there is a similar statement for $m^2\neq0$. This leads to a class of solutions given by
\begin{eqnarray}
V=\tilde V_0e^{(b-a)\phi} & = & -\frac{1}{2}\left(\theta^2-6\theta+8+2z^2-2\theta z+2z+\alpha^2\right)r^{-\theta}\,,
\label{eq:1aa} \\
A_t & = & \pm e^{-a\phi_0/2}\left(\frac{2(z-1)}{z+\tfrac{b\alpha}{2}}\right)^{1/2}r^{-z-b\alpha/2}\,,\\
m^2 & = & \frac{z+\tfrac{b\alpha}{2}}{2(z-1)}\left[(\theta-2)(\theta-2z+2)-\alpha^2\right]e^{(a-b)\phi_0}\,,\\
b & = & \frac{\alpha^2-\theta(\theta-2)}{(z-1)\alpha}\,, \label{eq:2aa} \\
a & = & \frac{\alpha^2-\theta(\theta-z-1)}{(z-1)\alpha}\,.\label{eq:am^2neq0}
\end{eqnarray}
Reality of $A_t$ requires that
\begin{equation}
\alpha^2\ge\theta(\theta-2)-2z(z-1)\,.
\end{equation}
The parameters of the solution are $\phi_0$, $\alpha$, $\theta$ and $z$. The sign of $m^2$ is not fixed by this reality condition and the NEC \eqref{eq:NEC1} and \eqref{eq:NEC2}. The parameters of the Lagrangian are $\tilde V_0$, $m^2$, $a$ and $b$. If we choose $\tfrac{b\alpha}{2}=2-\theta$, then the solution becomes the $m^2=0$ case given in \eqref{eq:phim=0case}--\eqref{eq:electricflux}. Under the Lifshitz scalings of \eqref{eq:scaling} we have that our 2-form field strength scales as $F\rightarrow\lambda^{-b\alpha/2}F$.

\subsection{Breaking of scale invariance}

For $\theta=0$ we obtain from \eqref{eq:1aa}--\eqref{eq:am^2neq0}
\begin{eqnarray}
\phi & = & \phi_0+\alpha\log r\,,\\
V & = & -z^2-z-4-\frac{1}{2}\alpha^2\,,\\
A_t & = & \pm e^{-a\phi_0/2}\left(\frac{2(z-1)}{z+\tfrac{b\alpha}{2}}\right)^{1/2}r^{-z-b\alpha/2}\,,\\
m^2 & = & \frac{z+\tfrac{b\alpha}{2}}{2(z-1)}\left[4(z-1)-\alpha^2\right]\,,\\
a=b & = & \frac{\alpha}{z-1}\,.
\end{eqnarray}
For $\alpha^2=4(z-1)$ we retrieve the Lifshitz solutions of the EMD model, \eqref{eq:Lifm^2=0_1}--\eqref{eq:Lifm^2=0_7}. It is in general not a regular limit to send $\alpha$ to zero. 

Since we have from \eqref{eq:2aa}, \eqref{eq:am^2neq0} that $(a-b)\alpha=\theta$ it follows that sending $\alpha$ to zero in \eqref{eq:phim^2neq0}--\eqref{eq:am^2neq0} is only possible if we also send $\theta$ to zero. Even though in general for the solutions above $\alpha$ is a free parameter (provided we allow the parameters of the Lagrangian to be free), we will always assume that $\alpha^2$ is some function of $\theta$ and $z$ admitting a Taylor expansion in $\theta$ around $\theta=0$. Assuming that we have
\begin{equation}\label{eq:x^2}
\alpha^2=c_1\theta+c_2\theta^2+\mathcal{O}(\theta^3)\,,
\end{equation}
where $c_1$, $c_2$, etc are functions of $z$, it can be shown that we need $c_1\neq 0$ in order for the limit $\theta\rightarrow 0$ to exist. In this case sending $\theta$ to zero leads to the Lifshitz solution \eqref{eq:Lifshitz1}--\eqref{eq:Lifshitz5} with $a=b=0$. 

In the $\theta=0$ case the space-time is Lifshitz and there is an additional Killing vector associated with the scaling \eqref{eq:scaling}. A natural requirement is to demand that for $\theta=0$ the matter fields respect this symmetry. This will be true if and only if $\alpha=0$, i.e. if the scalar $\phi$ is constant and the 1-form $A$ invariant. Breaking of the scale symmetry should then be controlled by $\theta$. These requirements tells us that we should take $\alpha$ as in \eqref{eq:x^2} and that the $\theta=0$ solution corresponds to $a=b=0$. In this way the breaking of the scale invariance for both the metric and all the matter fields is proportional to $\theta$. We can summarize this by saying that: ``$\theta$ needs a scalar field and $z$ needs a massive vector field''. Since in the massless case $\alpha$ is given by \eqref{eq:xm=0case} and $F$ scales as $F\rightarrow\lambda^{\theta-2}F$ this condition is not satisfied there. 

This is of course a choice. One could also demand that for $\theta$ going to zero we recover the Lifshitz solutions of the EMD model given by 
\eqref{eq:Lifm^2=0_1}--\eqref{eq:Lifm^2=0_7}. It would be interesting to understand the different choices better from the boundary perspective. 

In explicit models the Lagrangian parameters are fixed in which case we cannot continuously vary the $\theta$ parameter. In this case the above arguments do not apply and one must resort to another condition for how $\alpha^2$ depends on $\theta$ and $z$. In the case of the metric the fact that $ds^2\rightarrow \lambda^\theta ds^2$ under \eqref{eq:scaling} (and not for example $ds^2\rightarrow \lambda^{2\theta} ds^2$) follows from demandig that a black brane that is asymptotically hvLif has an entropy density that scales with temperature as $T^{(d-\theta)/z}$. Since the thermodynamics will not depend on the background scalar and vector fields one can use the EMD model to construct these black brane solutions \cite{Charmousis:2010zz,Huijse:2011ef} despite the fact that the background matter gives additional $\theta$ independent contributions to the breaking of scale invariance. It would be interesting to see if there exists a physical requirement that fixes $\alpha^2$ as a function of $\theta$ giving the coefficients $c_1$, $c_2$, etc in \eqref{eq:x^2} in terms of a universal, i.e. model independent, value possibly depending on $z$.

%%%%%%%%%%%%%%%%%%%%%%%%%%%%%%%%%%
%%%%%%%%%%%%%%%%%%%%%%%%%%%%%%%%%%
\subsection{Violating the WEC}\label{subsec:ViolatingWEC}

We have discussed in the previous subsection a model with a massive vector, defined by the EPD model  \eqref{eq:Lifmodel}, that allows for values of the exponents $\theta$ and $z$ satisfying the inequalities of \eqref{eq:NEC1} and \eqref{eq:NEC2}. The cases where one of these inequalities is saturated were specifically addressed. While the
 model \eqref{eq:Lifmodel} allows for values of $\theta$ and $z$ saturating the inequality \eqref{eq:NEC1}, this is not the case when we consider the inequality in \eqref{eq:NEC2}, provided $m^2 \geq 0$. More specifically, our model 
with $m^2 \geq 0$ fails to describe the following two NEC respecting cases
\begin{equation}
\theta=2\,,\qquad z<0\quad\text{or}\quad z>1\,,
\end{equation}
and 
\begin{equation}
\theta=2(z-1)\,,\qquad 1<z<4\,,
\end{equation}
One case of particular interest that belongs to the latter category has $\theta=1$ and hence $z=3/2$. This is the special case \eqref{thetaz} relevant to the description of non-Fermi liquids, which is a major motivation for the study of hvLif space-times. In this subsection, we will consider a model that generalizes \eqref{eq:Lifmodel}, and which allows for these exponents. This can be achieved either by taking $m^2<0$ or by adding an extra non-derivative term to the action of the form $e^{c\phi}A^4$ (see the end of this subsection).

Before presenting an explicit model, let us look at a general property that the matter fields must satisfy for $\theta=2$ and $\theta=2(z-1)$. Introduce the tetrad
\begin{equation}
e^{\underline{t}}_t=r^{\theta/2-z}\,,\qquad e^{\underline{r}}_r=e^{\underline{x}}_x=e^{\underline{y}}_y=r^{\theta/2-1}\,.
\end{equation}
We have
\begin{eqnarray}
T_{\underline{t}\underline{t}}+T_{\underline{r}\underline{r}} & = & \frac{1}{2}r^{-\theta}(\theta-2)(\theta+2-2z)\,,\\
T_{\underline{t}\underline{t}}+T_{\underline{x}\underline{x}} & = & r^{-\theta}(z-1)(z+2-\theta)\,,\\
T_{\underline{x}\underline{x}}+T_{\underline{r}\underline{r}} & = & r^{-\theta}(z+1-\theta)(z+2-\theta)\,,
\end{eqnarray}
and $T_{\underline{y}\underline{y}}=T_{\underline{x}\underline{x}}$. It follows that for $\theta=2(z-1)$ we have
\begin{equation}\label{eq:relationTcomponents}
T_{\underline{r}\underline{r}}=-T_{\underline{t}\underline{t}}=(2-z)T_{\underline{x}\underline{x}}=(2-z)T_{\underline{y}\underline{y}}=(z-4)(z-2)r^{-\theta}\,.
\end{equation}
We thus find
\begin{eqnarray}
T_{\underline{t}\underline{t}}\ge 0 & \rightarrow & 2\le z\le 4\qquad (2\le\theta\le 6)\,,\\
T_{\underline{t}\underline{t}}\le 0 & \rightarrow & 1\le z\le 2\qquad (0\le\theta\le 2)\,.\label{eq:WECviolated}
\end{eqnarray}
For $\theta=2$ we find 
\begin{eqnarray}
T_{\underline{t}\underline{t}} & = & 0\,,\\
T_{\underline{r}\underline{r}} & = & 0\,,\\
T_{\underline{x}\underline{x}} & = & (z-1)zr^{-2}\,.
\end{eqnarray}
We will focus on the case $\theta=2(z-1)$. We see from \eqref{eq:WECviolated} that the weak energy condition (WEC) is violated for $1\le z\le 2$. The includes $AdS$ in which the violation of the WEC comes from a constant negative potential, which is not problematic. From \eqref{eq:relationTcomponents} it is clear that violation of the WEC cannot (only) come from terms proportional to the metric in the energy-momentum tensor. We will therefore look for other non-derivative terms that can be added to the model that violate the WEC and that allow for NEC respecting $\theta=2(z-1)$ and $\theta=2$ solutions.

Let us consider again the solution \eqref{eq:phim^2neq0}--\eqref{eq:am^2neq0} of the previous subsection. Putting $\theta=2(z-1)$ (which includes $\theta=1$ and $z=3/2$) we obtain
\begin{eqnarray}
V=\tilde V_0e^{(b-a)\phi} & = & -\left((z-4)(z-3)+\frac{1}{2}\alpha^2\right)r^{-2(z-1)}\,,\\
A_t & = & \pm e^{-a\phi_0/2}\left(\frac{2(z-1)}{z+\tfrac{b\alpha}{2}}\right)^{1/2}r^{-z-b\alpha/2}\,,\\
m^2 & = & -\frac{z+\tfrac{b\alpha}{2}}{2(z-1)}\alpha^2 e^{(a-b)\phi_0}\,,\\
b & = & \frac{\alpha}{z-1}-\frac{4(z-2)}{\alpha}\,,\\
a & = & \frac{\alpha}{z-1}-\frac{2(z-3)}{\alpha}\,.
\end{eqnarray}
We observe that $m^2<0$ which together with the potential is responsible for the violation of the WEC. In the next section we will study the question of whether or not it is admissbale to let $m^2<0$. To this end we look at linearized perturbations around the background hvLif space-time. We will show that there are no immediate problems.

Instead of letting $m^2$ become negative we can also consider to add the following term $-\tfrac{\gamma}{4}e^{c\phi}A^4$ to the action \eqref{eq:Lifmodel}. In order to violate the WEC with $m^2>0$ we need $\gamma>0$. Note that in the energy-momentum tensor we have the additional term
\begin{equation}
+\gamma e^{c\phi}A^2\left(A_\mu A_\nu-\frac{1}{4}A^2g_{\mu\nu}\right)\,.
\end{equation}
Since we have on-shell $A^2<0$ it is for positive $\gamma$ that we can violate the WEC. We will consider this term again in section \ref{sec:AdSLifasymptotics}.

%%%%%%%%%%%%%%%%%%%%%%%%%%%%%%%%%%
%%%%%%%%%%%%%%%%%%%%%%%%%%%%%%%%%%

\section{Perturbations and probe fields}

In this section, we will study perturbations in the EDP model around the hvLif background given by \eqref{eq:phim^2neq0}--\eqref{eq:am^2neq0}, for general $z$ and $\theta$. Our main goal is to determine whether the purely radial perturbations (lowest lying perturbations) lead to any pathologies for negative values of $m^2<0$. We will show that this is not the case. As an application we derive from the perturbation analysis natural actions for probe fields and derive a BF bound for scalar and vector probes.

\subsection{Linearized perturbations}
\label{sec:linear}

We start by expanding the fields in the action \eqref{eq:Lifmodel},
\begin{eqnarray}
g_{\mu\nu} & = & \bar g_{\mu\nu}+\epsilon h_{\mu\nu}\,,\label{eq:metricpertb2}\\
A_\mu & = & \bar A_\mu+\epsilon \tilde a_\mu\,,\label{eq:vectorpertb2}\\
\phi & = & \bar\phi+\epsilon\varphi\,.\label{eq:scalarpertb2}
\end{eqnarray}
where the hvLif background (barred fields) is defined in \eqref{eq:phim^2neq0}--\eqref{eq:am^2neq0}. We write
\begin{eqnarray}
\bar A_t & = & A_0r^{-\kappa}\,,\qquad\kappa=z+\frac{b\alpha}{2}\,,\\
\bar\phi & = & \phi_0+\alpha\log r\,.
\end{eqnarray}
We will employ radial gauge throughout, i.e. we take 
\begin{equation}
h_{r\mu}=0\,.
\end{equation}
At first order in $\epsilon$, we get the linearized equations of motion:
\begin{eqnarray}
0 & = & \bar\nabla_\mu\bar\nabla_\nu h+\bar\square h_{\mu\nu}-\bar\nabla_\sigma\bar\nabla_\mu h_\nu{}^\sigma-\bar\nabla_\sigma\bar\nabla_\nu h_\mu{}^\sigma+\partial_\mu\bar\phi\partial_\nu\varphi+\partial_\nu\bar\phi\partial_\mu\varphi\nonumber\\
&&+m^2e^{b\bar\phi}\left(\bar A_\mu \tilde a_\nu+\bar A_\nu\tilde a_\mu+b\varphi\bar A_\mu\bar A_\nu\right)+\varphi V'(\bar\phi)\bar g_{\mu\nu}+V(\bar\phi)h_{\mu\nu}\nonumber\\
&&+e^{a\bar\phi}a\varphi\left(\bar F_{\mu\rho}\bar F_\nu{}^\rho-\frac{1}{4}\bar g_{\mu\nu}\bar F^2\right)+e^{a\bar\phi}\left(\bar F_{\mu\rho}\tilde f_\nu{}^\rho+\bar F_{\nu\rho}\tilde f_\mu{}^\rho-\bar F_{\mu\rho}\bar F_{\nu\sigma}h^{\rho\sigma}\right)\nonumber\\
&&-\frac{1}{2}e^{a\bar\phi}\bar g_{\mu\nu}\left(\bar F_{\rho\sigma}\tilde f^{\rho\sigma}-\bar F_{\tau\rho}\bar F^\tau{}_\sigma h^{\rho\sigma}\right)-\frac{1}{4}e^{a\bar\phi}\bar F^2h_{\mu\nu}\,,\label{eq:firstordermetric}\\
0 & = & \partial_\mu\left(\sqrt{-\bar g}e^{a\bar\phi}\left[a\varphi\bar F^{\mu\nu}+\tilde f^{\mu\nu}-\bar F^\mu{}_\rho h^{\nu\rho}+\bar F^\nu{}_\rho h^{\mu\rho}+\frac{1}{2}h\bar F^{\mu\nu}\right]\right)\nonumber\\
&&-m^2\sqrt{-\bar g}e^{b\bar\phi}\left[b\varphi\bar A^\nu+\tilde a^\nu-h^{\nu\rho}\bar A_\rho+\frac{1}{2}h\bar A^\nu\right]\,,\label{eq:firstordervector}\\
0 & = & \partial_\mu\left(\sqrt{-\bar g}\left[\bar g^{\mu\nu}\partial_\nu\varphi-h^{\mu\nu}\partial_\nu\bar\phi\right]+\frac{1}{2}\sqrt{-\bar g}h\bar g^{\mu\nu}\partial_\nu\bar\varphi\right)\nonumber\\
&&-\frac{a}{4}\sqrt{-\bar g}e^{a\bar\phi}\left(a\varphi\bar F^2+2\bar F_{\mu\nu}\tilde f^{\mu\nu}-2\bar F_{\mu\rho}\bar F^\mu{}_\sigma h^{\rho\sigma}+\frac{1}{2}h\bar F^2\right)\nonumber\\
&&-\frac{b}{2}m^2\sqrt{-\bar g}e^{b\bar\phi}\left(b\varphi\bar A^2+2\bar A_\mu\tilde a^\mu-h^{\mu\nu}\bar A_\mu\bar A_\nu+\frac{1}{2}h\bar A^2\right)\nonumber\\
&&-\sqrt{-\bar g}\varphi\frac{d^2V}{d\phi^2}(\bar\phi)-\frac{1}{2}\sqrt{-\bar g}h\frac{dV}{d\phi}(\bar\phi)\,,\label{eq:firstorderscalar}
\end{eqnarray}
where $\tilde f_{\mu\nu}=\partial_\mu\tilde a_\nu-\partial_\nu\tilde a_\mu$, $h=\bar g^{\mu\nu}h_{\mu\nu}$ and where $\bar\nabla$ is the covariant derivative with respect to the background metric $\bar g_{\mu\nu}$. Indices are raised/lowered with respect to $\bar g_{\mu\nu}$. We collect several equations in Appendix \ref{subsec1:appendix}.

We can use the symmetries of the background to expand in Fourier modes. Each mode is characterized by a frequency $\omega$ and a wave-vector $\vec k=(k_x,k_y)$,
\begin{eqnarray}
\varphi & = & e^{-i\omega t+i\vec k\cdot\vec x}g_1\,,\\
a_t & = & e^{-i\omega t+i\vec k\cdot\vec x}r^{-z}g_2\,,\\
a_r & = & e^{-i\omega t+i\vec k\cdot\vec x}r^{-1}g_3\,,\\
a_x & = & e^{-i\omega t+i\vec k\cdot\vec x}r^{-1}g_4\,,\\
a_y & = & e^{-i\omega t+i\vec k\cdot\vec x}r^{-1}g_5\,,\\
h^t{}_t & = & e^{-i\omega t+i\vec k\cdot\vec x}g_6\,,\\
h^x{}_t & = & e^{-i\omega t+i\vec k\cdot\vec x}r^{1-z}g_7\,,\\
h^y{}_t & = & e^{-i\omega t+i\vec k\cdot\vec x}r^{1-z}g_8\,,\\
h^x{}_x & = & e^{-i\omega t+i\vec k\cdot\vec x}g_9\,,\\
h^x{}_y & = & e^{-i\omega t+i\vec k\cdot\vec x}g_{10}\,,\\
h^y{}_y & = & e^{-i\omega t+i\vec k\cdot\vec x}g_{11}\,,
\end{eqnarray}
where we suppressed the $\omega$ and $\vec k$ indices on the complex fields $g_1$ to $g_{11}$. The perturbations $a_\mu$ are related to $\tilde a_\mu$ through $\tilde a_\mu= \bar A_t r^z a_\mu$. The resulting equations of motion are scale invariant under $r\rightarrow\lambda r$ if we also rescale $\vec k\rightarrow\lambda^{-1}\vec k$ and $\omega\rightarrow\lambda^{-z}\omega$. Notice that $h_{xy}$ and $h_{xx}-h_{yy}$ are tensor perturbations with respect to the isometry group of the $x-y$ plane. Likewise, $h_{ti}$ and $a_i$ with $i=x,y$ are vector perturbations; and finally $h_{xx}+h_{yy}$, $h_{tt}$, $\varphi$, $a_t$ and $a_r$ are scalar perturbations.

Imposing that the perturbations are real requires $g_i^{(\omega,\vec k)}$ to satisfy
\begin{equation}
g_i^{(\omega,\vec k)}{}^*=g_i^{(-\omega,-\vec k)}\,.
\end{equation}
If $g_i^{(\omega,\vec k)}$ is a solution, then it will satisfy this property since complex conjugating the equations and replacing $(\omega,\vec k)$ by $(-\omega,-\vec k)$ leaves the equations invariant. However, when $(\omega,\vec k)=0$ we need to ensure that it is consistent to have real perturbations $g_i$. To ensure that this is the case, we look now at the equations for $(\omega,\vec k)=0$, that is, we consider purely radial perturbations.

\subsection{Purely radial perturbations} 

For $(\omega,\vec k)=0$, the system of equations decouples into several sectors. We have two equations involving only $g_3$,
\begin{eqnarray}
0 & = &\kappa\left(\theta-2+\kappa-z\right)g_3\,, \label{eq:g3_1}\\
&&\nonumber\\
0 & = & -2(z-1)(\theta-2+\kappa-z)g_3\,,\label{eq:g3_2}
\end{eqnarray}
which enforce $g_3=0$ (otherwise we would have to take $m^2=0$). Two other perturbations completely decouple: $g_{10}$ and the combination $g_9-g_{11}$. They obey
\begin{eqnarray}
0 & = & r^2g_{10}''+(\theta-z-1)rg_{10}'\,,\label{eq:g10}\\
&&\nonumber\\
0 & = & r^2\left(g_{9}''-g_{11}''\right)+(\theta-z-1)r\left(g_{9}'-g_{11}'\right)\,,\label{eq:g9-g11}
\end{eqnarray}
respectively. These functions decouple because they correspond to $h_{xy}$ and $h_{xx}-h_{yy}$, which are tensor perturbations, as mentioned above. The two equations do not represent any restriction on the reality of $g_{10}$ and $g_9-g_{11}$.

The remaining equations split into three systems of equations that are mutually decoupled. We have a system for $g_4$ and $g_7$
(corresponding to vector perturbations $a_x$ and $h_{tx}$),
\begin{eqnarray}
0 & = & r^2g_4''+(\theta-z-1)rg_4'-\kappa rg_7'+(z-1)\left[\theta+2\kappa-2z-1\right]g_4+(z-1)\kappa g_7\,, \qquad\; \label{eq:subsyst1_1}\\
&&\nonumber\\
0 & = & r^2g_7''-2(z-1)rg_4'+(\theta-z-1)rg_7'-2(z-1)(\theta-3)g_4-(z-1)(\theta-3)g_7\,, \qquad\; \label{eq:subsyst1_2}
\end{eqnarray}
a system for $g_5$ and $g_8$ (corresponding to vector perturbations $a_y$ and $h_{ty}$),
\begin{eqnarray}
0 & = & r^2g_5''+(\theta-z-1)rg_5'-\kappa rg_8'+(z-1)\left[\theta+2\kappa-2z-1\right]g_5+(z-1)\kappa g_8\,,\qquad\;\label{eq:subsyst2_1}\\
&&\nonumber\\
0 & = & r^2g_8''-2(z-1)rg_5'+(\theta-z-1)rg_8'-2(z-1)(\theta-3)g_5-(z-1)(\theta-3)g_8\,,\qquad\;\label{eq:subsyst2_2}
\end{eqnarray}
and a system for the remaining (scalar) perturbations,
\begin{eqnarray}
0 & = & r^2g_1''+(\theta-z-1)rg_1'-2a(z-1)rg_2'+\frac{\alpha}{2}r\left(g_6'+g_9'+g_{11}'\right)\nonumber\\
&&-\left[(z-2)(z-1)ab-2(z-1)za^2-\theta(\theta-z-2)\right]g_1\nonumber\\
&&+(z-1)\left[2za-(\theta-4)b\right]\left(g_2-\frac{1}{2}g_6\right)\,,\label{eq:subsyst3_1}\\
&&\nonumber\\
0 & = & r^2g_2''-\kappa a rg_1'+(\theta-z-1)rg_2'+\frac{1}{2}\kappa r\left(g_6'-g_9'-g_{11}'\right)\nonumber\\
&&+(b-a)\kappa\left(\theta-2+\kappa-z\right)g_1\,,\label{eq:subsyst3_2}\\
&&\nonumber\\
0 & = & r^2g_6''+2(z-1)rg_2'+\frac{1}{2}(3\theta-4z-2)rg_6'+\frac{1}{2}(\theta-2z)r\left(g_9'+g_{11}'\right)\nonumber\\
&&+\left[(\theta-z-2)(\theta-2)(a-b)+2(z-1)(\theta-z-2)b+(z-1)b\kappa\right]g_1\nonumber\\
&&+2(z-1)(2\theta-4+\kappa-2z)\left(g_2-\frac{1}{2}g_6\right)\,,\label{eq:subsyst3_3}\\
&&\nonumber\\
0 & = & r^2(g_6''+g_9''+g_{11}'')+2\alpha rg_1'+2(z-1)rg_2'+\frac{1}{2}(\theta-4z+2)rg_6'\nonumber\\
&&+\frac{1}{2}(\theta-2)r(g_9'+g_{11}')+\left[(\theta-z-2)(\theta-2)(a-b)-b(z-1)\kappa\right]g_1\nonumber\\
&&-2(z-1)\kappa\left(g_2-\frac{1}{2}g_6\right)\,,\label{eq:subsyst3_4}
\end{eqnarray}

\begin{eqnarray}
0 & = & r^2\left(g_9''+g_{11}''\right)-4(z-1)rg_2'+(\theta-2)rg_6'+(2\theta-z-3)r\left(g_9'+g_{11}'\right)\nonumber\\
&&+\left[2(\theta-z-2)(\theta-2)(a-b)+2(2a-b)(z-1)\kappa\right]g_1 \nonumber\\
&&+4(z-1)\kappa \left(g_2-\frac{1}{2}g_6\right)\,.\label{eq:subsyst3_5}
\end{eqnarray}

We now solve \eqref{eq:subsyst1_1} and \eqref{eq:subsyst1_2} by writing it as a system of 4 first order equations. Defining
\begin{eqnarray}
h_1 & = & g_4\,,\\
h_2 & = & g_7\,,\\
h_3 & = & rg_4'\,,\\
h_4 & = & rg_7'\,,
\end{eqnarray}
the equations \eqref{eq:subsyst1_1} and \eqref{eq:subsyst1_2}  become
\begin{eqnarray}
rh_1' & = & h_3\,,\\
rh_2' & = & h_4\,,\\
rh_3' & = & -(\theta-z-2)h_3+\kappa h_4-(z-1)\left[\theta+2\kappa-2z-1\right]h_1-(z-1)\kappa h_2\,,\label{eq2:subsyst1_1}\\
&&\nonumber\\
rh_4' & = & 2(z-1)h_3-(\theta-z-2)h_4+2(z-1)(\theta-3)h_1+(z-1)(\theta-3)h_2\,,\label{eq2:subsyst1_2}
\end{eqnarray}
or as a matrix equation
\begin{equation}
r\frac{d}{dr}h_i=M_i{}^jh_j\,,
\end{equation}
where the $M_i{}^j$ are the components of the following matrix 
\begin{equation}
M=\left(\begin{array}{cccc}
0 & 0 & 1 & 0\\
0 & 0 & 0 & 1\\
-(z-1)\left[\theta+2\kappa-2z-1\right] & -(z-1)\kappa & -(\theta-z-2) & \kappa\\
2(z-1)(\theta-3) & (z-1)(\theta-3) & 2(z-1) & -(\theta-z-2)
\end{array}\right)\,.
\end{equation}
The subsystem of equations \eqref{eq:subsyst2_1} and \eqref{eq:subsyst2_2} is described by the same matrix $M$. The four eigenvalues of $M$ are all real and given by $3-\theta$, $1-z$, $z-1$ and $2z-\theta+1$. Hence the two subsystems \eqref{eq:subsyst1_1}--\eqref{eq:subsyst1_2} and \eqref{eq:subsyst2_1}--\eqref{eq:subsyst2_2} have real solutions (for real integration constants), and so there are no constraints coming from imposing reality.

Hence we are left with equations \eqref{eq:subsyst3_1}--\eqref{eq:subsyst3_5}. We can readily define a new function
\begin{equation}
\eta=\frac{g_9+g_{11}}{2}\,,
\end{equation}
so that the system of five equations depends only on the four functions $g_1$, $g_2$, $g_6$ and $\eta$. To be precise, the system does not depend on the function $\eta$ itself, but only on its derivatives. The five equations are not all independent. To identify a system of independent equations, we start by solving the five equations algebraically for $g_1''$, $g_2''$, $g_6''$, $\eta''$ and $\eta'$, in terms of $g_1'$, $g_1$, $g_2'$, $g_2$, $g_6'$ and $g_6$. The constraint is the equation for $\eta'$,
\begin{align}
\label{etaconst}
 &(1+z-\theta) \,r \eta' = -\frac{\alpha}{2} rg_1' + (1-z) rg_2' +\frac{\theta-2}{2} rg_6'  \nonumber \\
&\quad + \frac{\alpha^4+\alpha^2 \left[-2 (\theta -1) \theta +z^2+(2 \theta -3) z+2\right]+\theta  (-\theta +z+1) [2 (z-1) z-(\theta -2) \theta ]}{2 \alpha (z-1)} g_1  \nonumber \\
&\quad + [-\theta ^2+\theta +\alpha^2+z^2+(\theta -3) z+2]g_2 \nonumber \\
&\quad + \frac{1}{2} \left[\theta ^2-\alpha^2-z^2-\theta  (z+1)+3 z-2\right]g_6 \,.
\end{align}
We can now substitute in the remaining four equations the expression \eqref{etaconst} for $\eta'$ and the corresponding derivative, $\eta''$, obtained by differentiating \eqref{etaconst}. The consistency condition is that the four equations, after the substitution, are algebraically equivalent to three final equations for $g_1''$, $g_2''$ and $g_6''$, which are independent of $\eta$ and its derivatives. We will express these final equations in a matrix form as before,
\begin{equation}
r\frac{d}{dr}h_i=M_i{}^jh_j\,,
\end{equation}
defining
\begin{eqnarray}
h_1 & = & g_1\,,\\
h_2 & = & g_2\,,\\
h_3 & = & g_6\,,\\
h_4 & = & rg_1'\,,\\
h_5 & = & rg_2'\,,\\
h_6 & = & rg_6'\,.
\end{eqnarray}
The $M_i{}^j$ are the components of the following matrix 
\begin{equation}
M=\left(\begin{array}{cccccc}
0 & 0 & 0 & 1 & 0 & 0\\
0 & 0 & 0 & 0 & 1 & 0\\
0 & 0 & 0 & 0 & 0 & 1\\
M_4^{\phantom{s}1} & M_4^{\phantom{s}2} & M_4^{\phantom{s}3} & M_4^{\phantom{s}4} & M_4^{\phantom{s}5} & M_4^{\phantom{s}6} \\
M_5^{\phantom{s}1} & M_5^{\phantom{s}2} & M_5^{\phantom{s}3} & M_5^{\phantom{s}4} & M_5^{\phantom{s}5} & M_5^{\phantom{s}6} \\
M_6^{\phantom{s}1} & M_6^{\phantom{s}2} & M_6^{\phantom{s}3} & M_6^{\phantom{s}4} & M_6^{\phantom{s}5} & M_6^{\phantom{s}6} 
\end{array}\right)\,,
\label{eq:Mscalar}
\end{equation}
where the matrix components are listed in Appendix \ref{subsec2:appendix}. The six eigenvalues of this matrix are
\begin{align}
& 0 \,,\\
& 2+z-\theta \,,\\
& 2+z-\theta \,,\\
& -\frac{\theta }{2} \,,\\
& \frac{1}{2} \left(2+z-\theta +\frac{\sqrt{\sigma }}{\alpha (z-1)}\right) \,,\\
& \frac{1}{2} \left(2+z-\theta -\frac{\sqrt{\sigma }}{\alpha (z-1)}\right) \,,
\end{align}
with
\begin{eqnarray}
\sigma & = & 4 \alpha^6+4 \alpha^4 \left(-3 \theta ^2+3 z^2+2 \theta  (z+2)-7 z+4\right)+ \alpha^2 \left[12 \theta ^4-16 \theta ^3 (z+2)\right.  \nonumber \\
&& \left.+\theta ^2 ((62-11 z) z-3)+2 \theta  (z-1) \left(5 z^2+z-14\right)+(z-1)^2 (z (9 z-20)+20)\right] \nonumber\\
&& -4 (\theta -2) \theta ^2 (-\theta +z+1) \left(-\theta ^2+\theta +\theta  z+(z-3) z+2\right) \nonumber \\
\,.
\end{eqnarray}

For the special exponents
\begin{eqnarray}
\theta = 1\,, \qquad z=\frac{3}{2}\,,
\end{eqnarray}
we get real eigenvalues only, since
\begin{eqnarray}
\sigma = 4 \alpha^6+17 \alpha^4+\frac{361 \alpha^2}{16}+\frac{15}{2}>0\,.
\end{eqnarray}
Therefore, we do not expect a BF-type instability, which corresponds to a complex exponent. Likewise, for any pure Lifshitz space-time, $\theta=0$, we have positivity for $\alpha^2>0$,
\begin{eqnarray}
\sigma = 4 \alpha^2 \left( \left(\alpha^2+\frac{1}{2} (z-1) (3 z-4)\right)^2+(z+1) (z-1)^2 \right)\,.
\end{eqnarray}

For generic values of $z$ and $\theta$, assuming $z\geq 1$ and $\theta \geq 0$, the sign of $\sigma$ is undetermined. The physical conditions (area-law entanglement entropy and null energy condition) combined lead to
\begin{equation}
\label{eq:condsigma}
0\leq \theta \leq d-1=1\,,\qquad  z\geq 1 + \frac{\theta}{d} = 1 + \frac{\theta}2 \,.
\end{equation}
These are not enough to fix the sign of $\sigma$. To obtain a positive $\sigma$, and thus real eigenvalues, it is sufficient to further impose, along with those conditions, one of the following three restrictions:
\begin{equation}
z \geq 2\,,\qquad \theta \geq 0.10558 \,,\qquad \alpha^2 \geq 0.0058351 \,.
\end{equation}
There is a region with small values of $\theta$ and $\alpha^2$ for which $\sigma$ is negative (though not in the $\theta=0$ Lifshitz case).

In the case $\theta=d-1=1$ (for which the area law of the holographic entanglement entropy has a logarithmic violation), $\sigma$ is positive, as the comment above shows. However, if $\theta>d-1$, there is again a range for which $\sigma$ is negative.

%%%%%%%%%%%%%%%%%%%%%%%%%%%%%%%%%%%%%%%%%%%%%%%%%
%%%%%%%%%%%%%%%%%%%%%%%%%%%%%%%%%%%%%%%%%%%%%%%%%

\subsection{Probe fields and the BF bound}

We will now consider probe fields on the hvLif space-time given by \eqref{eq:phim^2neq0}--\eqref{eq:am^2neq0}, and discuss the analogue of the BF bound in AdS. We will choose particular interaction terms, which are the counterparts of a mass term in AdS, that have the property that the equations of motion for the probe fields are invariant under the Lifshitz scaling 
\begin{equation}
r \to \lambda r\,, \qquad x_i \to \lambda x_i\,, \qquad t \to \lambda^{z} t \,,
\end{equation}
(with an appropriate scaling of the vector components in the case of a probe vector field as we will see below). This symmetry is analogous to that of the linearized equations of motion studied in the previous section, and listed in  Appendix \ref{subsec1:appendix}.

\subsubsection*{Scalar probes}

Consider the following action for a probe scalar field $\chi$ on the hvLif background \eqref{eq:phim^2neq0}--\eqref{eq:am^2neq0},
\begin{equation}
S_\chi=\int d^4x\sqrt{-\bar g}\left(-\frac{1}{2}(\partial \chi)^2+c_R \,\bar R \chi^2 + c_F \,\frac{1}{4}e^{a\bar\phi}\bar F^2 \chi^2 +c_A \, \frac{1}{2}e^{b\bar\phi}\bar A^2 \chi^2 -c_\phi \,\frac{1}{2}e^{(b-a)\bar\phi} \chi^2\right)\,,
\end{equation}
where $c_R$, $c_F$, $c_A$ and $c_\phi$ are the coupling constants of the interactions of the scalar field. We will see what are the constraints on these couplings imposed by a BF-type bound in the hvLif space-time. The equation of motion for the scalar field is
\begin{eqnarray}
0 =  -r^{2z} \partial_t^2 \chi +r^2 \partial_x^2 \chi +r^2 \partial_y^2 \chi +r^2 \partial_r^2 \chi -(1+z-\theta) r \partial_r \chi -m^2_\chi \chi \,,
\end{eqnarray}
where we defined
\begin{equation}
m^2_\chi= c_R\, \left(3 (\theta -2)^2+4 z^2+(8-6 \theta ) z\right)+ c_F\, 2 (z-1)\kappa+ \left(c_A\,\frac{ 2 (z-1)}{\kappa } + c_\phi\right) e^{-\frac{\theta \phi_0}{\alpha}} \,.
\end{equation}
The choice of terms in the action guarantees that we obtain a homogeneous radial equation if $\chi=\chi(r)$. Keeping only the radial dependence, the solution to the equation of motion is
\begin{equation}
\chi(r) = C^+ \, r^{\frac{1}{2} \left(2+z-\theta + \sqrt{(2+z-\theta)^2+4m^2_\chi}\right)} 
+ C^- \, r^{\frac{1}{2} \left(2+z-\theta - \sqrt{(2+z-\theta)^2+4m^2_\chi}\right)} \,.
\end{equation}
The BF bound for the scalar field is therefore
\begin{equation}
m^2_\chi \geq -\frac{(2+z-\theta)^2}{4}\,.
\end{equation}
This is a direct generalization of the AdS result for $z=1$, $\theta=0$.

\subsubsection*{Vector probes}

Consider now the case of a vector field $W_\mu$ on the hvLif background. Let us take the action
\begin{eqnarray}
S_W=\int &d^4x& \sqrt{-\bar g}\Bigg(-\frac{1}{4}(dW)^2+q_R \,\bar R\, W^2+q_{RW} \,\bar R_{\mu\nu} W^\mu W^\nu
\nonumber \\
&& + q_F \,\frac{1}{4}e^{a\bar\phi}\bar F^2 W^2 +q_A \, \frac{1}{2}e^{b\bar\phi}\bar A^2 W^2 +q_{AW} \, \frac{1}{2}e^{b\bar\phi}(\bar A_\mu W^\mu)^2
\nonumber \\
&&- q_{D} \,\frac{1}{2}(\partial \bar\phi)^2 W^2 - q_{DW} \,\frac{1}{2}((\partial_\mu \bar\phi) W^\mu)^2 -q_\phi \,\frac{1}{2}e^{(b-a)\bar\phi} W^2\Bigg)\,,
\end{eqnarray}
where $q_R$, $q_{RW}$, $q_F$, $q_A$, $q_{AW}$, $q_D$, $q_{DW}$ and $q_\phi$ are the coupling constants. We wrote down interaction terms which will lead to homogeneous radial dependence in the equations of motion for $W_\mu$, as happened in the case of the scalar field. The equations of motion are
\begin{eqnarray}
0 &=& r^{2} (\partial_r^2 W_t+\partial_x^2 W_t+\partial_y^2 W_t - \partial_t \partial_r W_r - \partial_t \partial_x W_x - \partial_t \partial_y W_y) \nonumber \\
&& +(z-1) r (\partial_r W_t-\partial_t W_r)  -m^2_{W_t} W_t \,, \\
&&\nonumber\\
0 &=& -r^{2z} (\partial_t^2 W_r - \partial_t \partial_r W_t) + r^{2} (\partial_x^2 W_r + \partial_y^2 W_r - \partial_r \partial_x W_x-  \partial_r \partial_y W_y) \nonumber \\
&& -m^2_{W_r} W_r \,, \\
&&\nonumber\\
0 &=& -r^{2z} (\partial_t^2 W_x - \partial_t \partial_x W_t) + r^{2} (\partial_r^2 W_x + \partial_y^2 W_x - \partial_r \partial_x W_r-  \partial_y \partial_x W_y) \nonumber \\
&& -(z-1) r (\partial_r W_x-\partial_x W_r)  -m^2_{W_x} W_x \,, \\
&&\nonumber\\
0 &=& -r^{2z} (\partial_t^2 W_y - \partial_t \partial_y W_t) + r^{2} (\partial_r^2 W_y + \partial_x^2 W_y - \partial_r \partial_y W_r-  \partial_x \partial_y W_x) \nonumber \\
&& -(z-1) r (\partial_r W_y-\partial_y W_r)  -m^2_{W_y} W_y \,,
\end{eqnarray}
where we defined
\begin{eqnarray}
m^2_{W_t} &=& q_R\, \left(3 (\theta -2)^2+4 z^2+(8-6 \theta ) z\right) +q_{RW} \,\left( (\theta -2) \theta +2 z^2+(4-3 \theta ) z \right) \nonumber \\
&& + q_F\, 2 (z-1)\kappa+ \left((2 q_A + q_{AW}) \,\frac{z-1}{\kappa } + q_\phi\right) e^{-\frac{\theta \phi_0}{\alpha}} + q_D \alpha^2 \,, \\
&&\nonumber\\
m^2_{W_r} &=& q_R\, \left(3 (\theta -2)^2+4 z^2+(8-6 \theta ) z\right) +q_{RW}\, \left(-2 \theta +2 z^2-\theta  z+4 \right) \nonumber \\
&& + q_F\, 2 (z-1)\kappa+ \left(q_A \,\frac{2(z-1)}{\kappa } + q_\phi\right) e^{-\frac{\theta \phi_0}{\alpha}} + (q_D+q_{DW}) \alpha^2 \,, \\
&&\nonumber\\
m^2_{W_x} &=& q_R\, \left(3 (\theta -2)^2+4 z^2+(8-6 \theta ) z\right) +q_{RW}\, (2-\theta)(2+z-\theta) \nonumber \\
&& + q_F\, 2 (z-1)\kappa+ \left(q_A \,\frac{2(z-1)}{\kappa } + q_\phi\right) e^{-\frac{\theta \phi_0}{\alpha}} + q_D\alpha^2 \,, \\
&&\nonumber\\
m^2_{W_y} &=& m^2_{W_x} \,.
\end{eqnarray}
Let us take the purely radial case, $W_\mu=W_\mu(r)$. Then the second equation above enforces $W_r=0$ (unless $m_{W_r}^2$=0). The other three equations decouple and have solutions
\begin{eqnarray}
W_t(r) &=& C^+_t \, r^{\frac{1}{2} \left(2-z + \sqrt{(2-z)^2+4m^2_{W_t}}\right)} 
+ C^-_t \, r^{\frac{1}{2} \left(2-z - \sqrt{(2-z)^2+4m^2_{W_t}}\right)}  \,, \nonumber \\
W_i(r) &=& C^+_i \, r^{\frac{1}{2} \left(z + \sqrt{z^2+4m^2_{W_i}}\right)} 
+ C^-_i \, r^{\frac{1}{2} \left(z - \sqrt{z^2+4m^2_{W_i}}\right)}  \,,
\end{eqnarray}
for $i=x,y$. Therefore, the BF bound for the vector field is the combination of the two conditions
\begin{equation}
m^2_{W_t} \geq -\frac{(2-z)^2}{4}\,,  \qquad m^2_{W_i} \geq -\frac{z^2}{4}\,.
\end{equation}

The standard AdS BF bound is determined near the boundary ($r=0$), where the non-radial derivatives are subleading in the equations of motion. In our case, although the terms would still be subleading, the boundary of the hvLif space-time is singular, due to the divergent behavior of the curvature invariants. Taking the perspective of \cite{Dong:2012se} that the hvLif space-time is an effective description for $r$ not too small nor too large, the small $r$ region requires an AdS or Lifshitz UV completion. However, the analysis above is still relevant, since all perturbation modes of the probe fields (and of the background fields studied before) should be real in the entire space-time, including the range described approximately by the hvLif metric. This is not possible unless the BF bounds determined above are respected.

It will be interesting to compute the two-point function for these probe fields.

%%%%%%%%%%%%%%%%%%%%%%%%%%%%%%%%%%
%%%%%%%%%%%%%%%%%%%%%%%%%%%%%%%%%%
\section{AdS or Lifshitz asymptotics}\label{sec:AdSLifasymptotics}

Curvature invariants of the hvLif metric blow up near the boundary at $r=0$ for $\theta>0$, so we should replace the UV with a regular $\theta=0$ space-time, i.e. AdS or Lifshitz. 
As already explained at the end of the introduction we will therefore consider in this section a class of models, i.e. Lagrangians with the same parameters, that admit both hvLif space-times as well as AdS or Lifshitz space-times. This is a useful starting point for the explicit construction of interpolating solutions with an AdS or Lifshitz UV and a hvLif IR. Generically the IR geometries are also not regular, so one could consider to further deform it. We will not consider this here.

A physically very interesting case is to put a $\theta=1$ geometry in the IR because this leads to logarithmic violations of the area law for entanglement entropy. The previous model \eqref{eq:Lifmodel} is too constrained for this purpose. This is because demanding that the equations of motion of \eqref{eq:Lifmodel} admit AdS solutions requires the potential to have an extremum. On the other hand, the potential in that case must be a single exponential in order to allow for hvLif solutions. So an AdS UV is not possible. For Lifshitz solutions, one does not need to sit at the extremum of a potential, but in that case, if demands that there are both $\theta=1$ and $\theta=0$ solutions there are conflicting conditions for the mass parameter. This situation gets much better if we consider multiple dilaton fields. However this is generically not good enough since for example an IR geometry with $\theta=1$ and $z=3/2$ requires a negative $m^2$ whereas a Lifshitz UV requires a positive $m^2$. This problem can be overcome by adding the $A^4$ term mentioned in Subsection \ref{subsec:ViolatingWEC} where we discussed ways of violating the WEC with non-derivative terms.

\subsection{Model and equations of motion}

So we now consider the following more general model
\begin{eqnarray}
S&=&\int d^4x\sqrt{-g}\left(R-\frac{1}{2}(\partial\vec\phi)^2-\frac{1}{4}e^{\vec a\cdot\vec\phi}F^2-\frac{m^2}{2}e^{\vec b\cdot\vec\phi}A^2-\frac{\gamma}{4}e^{\vec c\cdot\vec\phi}A^4-V(\vec\phi)\right)\,.\label{eq:Lifmodelmulti}
\end{eqnarray}
We will demand that this theory admits either an AdS or Lifshitz solution as well as a $\theta=1$ solution:
\begin{eqnarray}
ds^2 & = & r\left(-\frac{dt^2}{r^{2z}}+\frac{1}{r^2}\left(dr^2+dx^2+dy^2\right)\right)\,,\label{eq:IRmetric}\\
A & = & A_0r^{-\kappa}dt\,,\\
\vec\phi & = & \vec\phi_0+\vec \alpha\log r\,.\label{eq:IRscalar}
\end{eqnarray}
The latter is chosen for explicitness and for its physical relevance. 

The equations of motion for \eqref{eq:Lifmodelmulti} are
\begin{eqnarray}
G_{\mu\nu} & = & \frac{1}{2}\partial_\mu\vec\phi\cdot\partial_\nu\vec\phi-\frac{1}{4}g_{\mu\nu}\partial_\rho\vec\phi\cdot\partial^\rho\vec\phi+\frac{1}{2}e^{\vec a\cdot\vec\phi}\left(F_{\mu\rho}F_\nu{}^\rho-\frac{1}{4}g_{\mu\nu}F^2\right)-\frac{1}{2}Vg_{\mu\nu}\,,\nonumber\\
&&+\frac{1}{2}m^2e^{\vec b\cdot\vec\phi}\left(A_\mu A_\nu-\frac{1}{2}A^2g_{\mu\nu}\right)+\frac{1}{2}\gamma e^{\vec c\cdot\vec\phi}A^2\left(A_\mu A_\nu-\frac{1}{4}A^2g_{\mu\nu}\right)\,,\\
\nabla_\mu\left(e^{\vec a\cdot\vec\phi}F^{\mu\nu}\right) & = & m^2 e^{\vec b\cdot\vec\phi}A^\nu+\gamma e^{\vec c\cdot\vec\phi}A^2A^\nu\,,\\
\square\phi_i & = & \frac{a_i}{4}e^{\vec a\cdot\vec\phi}F^2+\frac{b_i}{2}m^2e^{\vec b\cdot\vec\phi}A^2+\frac{c_i}{4}\gamma e^{\vec c\cdot\vec\phi}A^4+\frac{\partial V}{\partial\phi_i}\,.
\end{eqnarray}

\subsection{Solutions}\label{subsec:solutions}

We define
\begin{eqnarray}
V(\vec\phi=\vec\phi_0+\vec \alpha\log r) & = & V_0r^{-\theta}\,,\\
\frac{\partial V}{\partial\phi_i}(\vec\phi=\vec\phi_0+\vec \alpha\log r) & = & \beta_ir^{-\theta}\,.
\end{eqnarray}
Substituting the Ansatz \eqref{eq:IRmetric}--\eqref{eq:IRscalar} (for the moment we keep $\theta$ arbitrary) with $\vec \alpha\neq 0$ into the equations of motion we find
\begin{eqnarray}
\vec a\cdot\vec \alpha & = & \theta-2z+2\kappa\,,\label{eq:adotx}\\
\vec b\cdot\vec \alpha & = & -2z+2\kappa\,,\label{eq:bdotx}\\
\vec c\cdot\vec \alpha & = & \theta-4z+4\kappa\,,\label{eq:cdotx}
\end{eqnarray}
as well as
\begin{eqnarray}
\vec \alpha^2+\tilde m^2\tilde A_0^2-\tilde\gamma\tilde A_0^4 & = & (\theta-2)(\theta+2-2z)\,,\\
\kappa^2\tilde A_0^2+\tilde m^2\tilde A_0^2-\tilde\gamma\tilde A_0^4 & = & 2(z-1)(z+2-\theta)\,,\\
2\tilde m^2\tilde A_0^2-\tilde\gamma\tilde A_0^4-4V_0 & = & 4(z+1-\theta)(z+2-\theta)\,,\\
\left(\kappa\tilde A_0^2-2(z-1)\right)(z+2-\theta) & = & 0\,,\\
\alpha_i(\theta-z-2) & = & -\frac{a_i}{2}\kappa^2\tilde A_0^2-\frac{b_i}{2}\tilde m^2\tilde A_0^2+\frac{c_i}{4}\tilde\gamma\tilde A_0^4+\beta_i\,,
\end{eqnarray}
where we defined
\begin{eqnarray}
\tilde A_0^2 & = & A_0^2e^{\vec a\cdot\vec\phi_0}\,,\\
\tilde m^2 & = & m^2e^{(\vec b-\vec a)\cdot\vec\phi_0}\,,\\
\tilde\gamma & = & \gamma e^{(\vec c-2\vec a)\cdot\vec\phi_0}\,.
\end{eqnarray}
Assuming $\theta\neq z+2$ this can be written as
\begin{eqnarray}
m^2 & = & \frac{\kappa}{z-1}\left((z+3-2\theta)(z+2-\theta)+(z-1)\kappa+2V_0\right)e^{(\vec a-\vec b)\cdot\vec\phi_0}\,,\\
\gamma & = & \frac{\kappa^2}{(z-1)^2}\left((\theta-2)(\theta-z-2)+(z-1)\kappa+V_0\right)e^{(2\vec a-\vec c)\cdot\vec\phi_0}\,,\\
\vec \alpha^2 & = & \theta(\theta-2)-2z(z-1)+2(z-1)\kappa\,,\\
A_0^2 & = & \frac{2}{\kappa}(z-1)e^{-\vec a\cdot\vec\phi_0}\,,\\
0 & = & \left[(z-1)\kappa+V_0+(\theta-z-2)(\theta-2)\right](\vec a+\vec b-\vec c)\nonumber\\
&&+\left[V_0+(\theta-z-2)(\theta-2)\right]\left(\vec b-\vec a-\frac{1}{V_0}\vec \beta\right)\nonumber\\
&&+(\theta-z-2)\left(\vec \alpha-(z-1)\vec b+(\theta-2)\frac{1}{V_0}\vec \beta\right)\,.
\end{eqnarray}
The linear combination of vectors in the last expression is such that each of them is orthogonal to $\vec \alpha$ as follows from \eqref{eq:adotx}--\eqref{eq:cdotx} as well as 
the fact that
\begin{equation}
\vec \alpha\cdot\vec \beta = -\theta V_0\,.
\end{equation}

If we instead consider an Ansatz of the form 
\begin{eqnarray}
ds^2 & = & -\frac{dt^2}{r^{2z}}+\frac{1}{r^2}\left(dr^2+dx^2+dy^2\right)\,,\\
A & = & A_0r^{-z}dt\,,\\
\vec\phi & = & \vec\phi_0\,,
\end{eqnarray}
i.e. with $\theta=0$ and matter respecting the Lifshitz scaling symmetry we obtain for $z>1$
\begin{eqnarray}
m^2 & = & \frac{2z}{z-1}\left(V_0+z^2+2z+3\right)e^{(\vec a-\vec b)\cdot\vec\phi_0}\,,\\
\gamma & = & \frac{z^2}{(z-1)^2}\left(V_0+z^2+z+4\right)e^{(2\vec a-\vec c)\cdot\vec\phi_0}\,,\\
A_0^2 & = & \frac{2}{z}(z-1)e^{-\vec a\cdot\vec\phi_0}\,,\\
0 & = & -z(z-1)\vec a-\left(2V_0+2z^2+4z+6\right)\vec b+\left(V_0+z^2+z+4\right)\vec c+\vec \beta\,,
\end{eqnarray}
where 
\begin{eqnarray}
V_0 & = & V(\vec\phi=\vec\phi_0)\,,\\
\beta_i & = & \frac{\partial V}{\partial\phi_i}(\vec\phi=\vec\phi_0)\,.
\end{eqnarray}
For $z=1$ we find 
\begin{eqnarray}
A_0 & = & 0\,,\\
V_0 & = & -6\,,\\
\vec \beta & = & 0\,,
\end{eqnarray}
with $m$ and $\gamma$ arbitrary. We see that for AdS we need an extremum of the potential $\beta_i=0$ whereas for Lifshitz we can have $\beta_i\neq 0$.

\subsubsection*{$\theta=1$ solutions}

In order that the equations of motion of \eqref{eq:Lifmodelmulti} admit $\theta=1$ solutions we need $\vec \alpha\neq 0$ and
\begin{eqnarray}
\vec \alpha^2 & = & -1+2(z-1)(\kappa-z)\,,\label{eq:alphax^2}\\
V_0 & = & rV\left(\vec\phi=\vec\phi_0+\vec \alpha\log r\right) \,,\\
\beta_i & = & r\frac{\partial V}{\partial\phi_i}(\vec\phi=\vec\phi_0+\vec \alpha\log r)\,,\\
\vec \beta & = & (z+1+(z-1)\kappa+V_0)(\vec a+\vec b-\vec c)+(V_0+z+1)(\vec b-\vec a)\nonumber\\
&&-(z+1)(\vec \alpha-(z-1)\vec b)\,,\label{eq:dilatoncouplingsIR}\\
m^2 & = & \frac{\kappa}{z-1}\left((z+1)^2+(z-1)\kappa+2V_0\right)e^{(\vec a-\vec b)\cdot\vec\phi_0}\,,\\
\gamma & = & \frac{\kappa^2}{(z-1)^2}\left(z+1+(z-1)\kappa+V_0\right)e^{(2\vec a-\vec c)\cdot\vec\phi_0}\,,\label{eq:gammaIR}
\end{eqnarray}
where 
\begin{eqnarray}
-V_0 & = & \vec \alpha\cdot\vec \beta \,,\\
0 & = & \vec \alpha\cdot\left(\vec c-\vec a-\vec b\right)\,,\label{eq:conditionabc}\\
1 & = & \vec \alpha\cdot\left(\vec a-\vec b\right)\,,\label{eq:conditionab}\\
1 & = & \vec \alpha\cdot\left((z-1)\vec b-\vec \alpha\right)\,.\label{eq:conditionb}
\end{eqnarray}

\subsubsection*{AdS solutions} 

In order that the model \eqref{eq:Lifmodelmulti} admits AdS solutions we need that 
\begin{eqnarray}
V(\vec\phi=\vec\phi_{\text{UV}}) & = & -6\label{eq:AdSUV1}\,,\\
\frac{\partial V}{\partial\phi_i}(\vec\phi=\vec\phi_{\text{UV}}) & = & 0\,,\label{eq:AdSUV2}
\end{eqnarray}
where by $\vec\phi_{\text{UV}}$ we mean the near boundary value of $\vec\phi$. Since for $z=1$ we have that the vector field $A$ vanishes the values of $m^2$, $\gamma$, $\vec a$, $\vec b$ and $\vec c$ are arbitrary.

\subsubsection*{Lifshitz solutions}

In order that the model \eqref{eq:Lifmodelmulti} admits Lifshitz solutions with $z_{\text{UV}}=Z>1$, i.e. solutions of the form
\begin{eqnarray}
ds^2 & = & -\frac{dt^2}{r^{2Z}}+\frac{1}{r^2}\left(dr^2+dx^2+dy^2\right)\,,\label{eq:UVmetric}\\
A & = & \pm\left(\frac{2(Z-1)}{Z}\right)^{1/2}e^{-\vec a\cdot\vec\phi_{\text{UV}}/2}r^{-Z}dt\,,\\
\vec\phi & = & \vec\phi_{\text{UV}}\,,\label{eq:UVscalar}
\end{eqnarray}
we need 
\begin{eqnarray}
m^2 & = & \frac{2Z}{Z-1}\left(Z^2+2Z+3+V_{\text{UV}}\right)e^{(\vec a-\vec b)\cdot\vec\phi_{\text{UV}}}\,,\label{eq:m^2UV}\\
\gamma & = & \frac{Z^2}{(Z-1)^2}\left(Z^2+Z+4+V_{\text{UV}}\right)e^{(2\vec a-\vec c)\cdot\vec\phi_{\text{UV}}}\,,\label{eq:gammaUV}
\end{eqnarray}
as well as
\begin{eqnarray}
\frac{\partial V}{\partial\phi_i}(\vec\phi=\vec\phi_{\text{UV}}) & = & V_{\text{UV}}(b_i-a_i)+(Z^2+Z+4+V_{\text{UV}})(a_i+b_i-c_i)\nonumber\\
&&-2(Z+2)a_i+(Z+2)(Z+1)b_i\,,\label{eq:conditionfirstderV_UV}
\end{eqnarray}
where
\begin{equation}
V_{\text{UV}}=V(\vec\phi=\vec\phi_{\text{UV}})\,.
\end{equation}

\subsection{Constant potential}

Consider the case where the potential $V=V_0$ is constant. The NEC restricts us to take in the UV $Z\ge 1$ and in the IR $z\ge 3/2$. We first consider the case where $Z>1$, i.e. we consider a Lifshitz UV. We then find for $m^2$ and $\gamma$
\begin{eqnarray}
m^2 & = & \frac{\kappa}{z-1}\left((z+1)^2+(z-1)\kappa+2V_0\right)e^{(\vec a-\vec b)\cdot\vec\phi_0}\nonumber\\
& = & \frac{2Z}{Z-1}\left(Z^2+2Z+3+V_0\right)e^{(\vec a-\vec b)\cdot\vec\phi_{\text{UV}}}\,,\label{eq:m2cstpot}\\
\gamma & = & \frac{\kappa^2}{(z-1)^2}\left(z+1+(z-1)\kappa+V_0\right)e^{(2\vec a-\vec c)\cdot\vec\phi_0}\nonumber\\
& = & \frac{Z^2}{(Z-1)^2}(Z^2+Z+4+V_0)e^{(2\vec a-\vec c)\cdot\vec\phi_{\text{UV}}}\,.\label{eq:gammacstpot}
\end{eqnarray}
For simplicity we will assume that
\begin{equation}
(\vec a+\vec b-\vec c)\cdot\left(\vec\phi_0-\vec\phi_{\text{UV}}\right)=0\,.
\end{equation}
We define
\begin{equation}
f=\frac{Z}{2(Z-1)}\frac{Z^2+Z+4+V_0}{Z^2+2Z+3+V_0}\,.
\end{equation}
From \eqref{eq:m2cstpot} and \eqref{eq:gammacstpot} it follows that
\begin{eqnarray}
0 & = & \kappa^2+\left(\frac{z+1+V_0}{z-1}-(z-1)f\right)\kappa-\left((z+1)^2+2V_0\right)f\,,\label{eq:alphacstpot}\\
e^{(\vec a-\vec b)\cdot\left(\vec\phi_0-\vec\phi_{\text{UV}}\right)} & = & \frac{Z^2}{(Z-1)^2}\frac{(z-1)^2}{\kappa^2}\frac{Z^2+Z+4+V_0}{z+1+(z-1)\kappa+V_0}>0\,.\label{eq:pos}
\end{eqnarray}
In order that $\kappa$ satisfying \eqref{eq:alphacstpot} is real we need that the discriminant $D$ given by
\begin{equation}\label{eq:D}
D=(z-1)^2f^2+2\left((z+1)(2z+1)+3V_0\right)f+\frac{(z+1+V_0)^2}{(z-1)^2}\ge 0\,.
\end{equation}
On top of that we need that 
\begin{equation}\label{eq:x2}
\vec \alpha^2=-1+2(z-1)(\kappa-z)>0\,.
\end{equation}
By appropriately choosing $V_0$ we can satisfy the positivity requirements \eqref{eq:pos}, \eqref{eq:D} and \eqref{eq:x2} for a wide range of values for $z\ge 3/2$ and $Z>1$. For example $z=3/2$, $Z=2$ and $V_0<V_0^*$ where $V_0^*\approx -11.18$ is the smallest real zero of $D$ and where $\kappa$ is the largest positive real number satisfying \eqref{eq:alphacstpot}, satisfies all positivity conditions \eqref{eq:pos}, \eqref{eq:D} and \eqref{eq:x2}. There are no serious constraints coming from the equations involving the dilaton parameters. In the IR we can read \eqref{eq:dilatoncouplingsIR} with $\vec \beta=0$ as providing $\vec \alpha$ whereas in the UV \eqref{eq:conditionfirstderV_UV} gives just a constraint on $\vec a$, $\vec b$ and $\vec c$.

Consider next an AdS UV, i.e. we set $Z=1$. In this case since the vector field goes to zero and the scalar becomes constant in the UV the values of $m^2$ and $\gamma$ are only dictated by the IR solution. In the case of a constant potential supporting AdS$_4$ we need to take $V_0=-6$. This gives
\begin{eqnarray}
m^2 & = & \frac{\kappa}{z-1}\left((z+1)^2+(z-1)\kappa-12\right)e^{(\vec a-\vec b)\cdot\vec\phi_0}\,,\\
\gamma & = & \frac{\kappa^2}{(z-1)^2}\left(z+1+(z-1)\kappa-6\right)e^{(2\vec a-\vec c)\cdot\vec\phi_0}\,,
\end{eqnarray}
with $z\ge 3/2$ where $\kappa$ can be taken to be any real number such that $\vec \alpha^2>0$. Further the AdS UV does not constrain the dilaton couplings $\vec a$, $\vec b$ and $\vec c$. Again \eqref{eq:dilatoncouplingsIR} with $\vec \beta=0$ can be viewed as giving $\vec \alpha$.

\subsection{Multi-exponential potentials} 

We can replace the constant potential by one depending on the scalars $\vec\phi$. A typical supergravity inspired choice would be to take a potential of the form
\begin{equation}
V=\sum_{I=1}^M\Lambda_Ie^{\vec\alpha_I\cdot\vec\phi}\,,
\end{equation}
which then needs to be constrained such that we can have both $\theta=1$ and $\theta=0$ solutions of the equations of motion of the action \eqref{eq:Lifmodelmulti}. Here we will just indicate the part of the analysis that is universal to all such potentials. We leave it for future applications to work out interesting special cases where one has, say, $M=1$, $M=2$, or $M=3$ exponential terms.

We have in the IR
\begin{eqnarray}
V(\vec\phi=\vec\phi_0+\vec \alpha\log r) & = & \sum_I\Lambda_Ie^{\vec\alpha_I\cdot\vec\phi_0}r^{\vec\alpha_I\cdot\vec \alpha}=V_0r^{-1}\,,\label{eq:potcondition1}\\
\frac{\partial V}{\partial\phi_i}(\vec\phi=\vec\phi_0+\vec \alpha\log r) & = & \sum_I\Lambda_Ie^{\vec\alpha_I\cdot\vec\phi_0}\alpha_{Ii}r^{\vec\alpha_I\cdot\vec \alpha}=\beta_ir^{-1}\,.\label{eq:potcondition2}
\end{eqnarray}
By contracting \eqref{eq:potcondition2} with $\alpha_i$ and adding \eqref{eq:potcondition1} using $\vec \alpha\cdot\vec \beta=-V_0$ we find
\begin{equation}
\sum_I\Lambda_Ie^{\vec\alpha_I\cdot\vec\phi_0}(1+\alpha_I\cdot\vec \alpha)r^{\vec\alpha_I\cdot\vec \alpha}=0\,.\label{eq:conditionLambdas}
\end{equation}
We split the sum over $I$ into three pieces
\begin{enumerate}
\item[$\bar I$]: those I for which $\alpha_I\cdot\vec \alpha=-1$,
\item[$\hat I$]: those I for which $\alpha_I\cdot\vec \alpha\neq-1$ and the value $\alpha_I\cdot\vec \alpha$ occurs only once,
\item[$\tilde I_a$]: those I for which $\alpha_I\cdot\vec \alpha\neq-1$ and the value $\alpha_I\cdot\vec \alpha$ occurs at least twice. A group $\tilde I_a$ contains all cases where $\alpha_I\cdot\vec \alpha\neq-1$ gives the same value.
\end{enumerate}
We illustrate this with an example. Let $\alpha_I\cdot\vec \alpha=(2,-1,1,-1,1,5,2,-1)$ where $I=1,\ldots,8$ then $\bar I=2,4,8$, $\hat I=6$, $\tilde I_1=1,7$ and $\tilde I_2=3,5$. The conditon \eqref{eq:conditionLambdas} gives no constraints for the $\tilde\Lambda_{\bar I}$. Further we find
\begin{eqnarray}
0 & = & \Lambda_{\hat I}\,,\\
0 & = & \sum_{\tilde I_a}\Lambda_{\tilde I_a}e^{\vec\alpha_{\tilde I_a}\cdot\vec\phi_0}\,.\label{eq:conditionLambdasIa}
\end{eqnarray}
Because of $\Lambda_{\hat I}=0$ we can assume WLOG from the start that there are no $\hat I$ cases. It follows also that
\begin{equation}
V_0=\sum_{\bar I}\Lambda_{\bar I}e^{\vec\alpha_{\bar I}\cdot\vec\phi_0}\,.
\end{equation}
Equation \eqref{eq:potcondition2} becomes
\begin{eqnarray}
\sum_{\bar I}\Lambda_{\bar I}e^{\vec\alpha_{\bar I}\cdot\vec\phi_0}\alpha_{\bar I i} & = & \beta_i\,,\\
\sum_{\tilde I_a}\Lambda_{\tilde I_a}e^{\vec\alpha_{\tilde I_a}\cdot\vec\phi_0}\alpha_{\tilde I_ai} & = & 0\,.\label{eq:conditionalphasIa}
\end{eqnarray}
Note that even though for the IR solution the indices $\tilde I_a$ play no role for the values of $V_0$ and $\beta_i$, this is not so for the UV solution where we have
\begin{equation}
V_{\text{UV}}=\sum_{\bar I}\Lambda_{\bar I}e^{\vec\alpha_{\bar I}\cdot\vec\phi_{\text{UV}}}+\sum_a\sum_{\tilde I_a}\Lambda_{\tilde I_a}e^{\vec\alpha_{\tilde I_a}\cdot\vec\phi_{\text{UV}}}\,.
\end{equation}
\begin{equation}
\frac{\partial V}{\partial\phi_i}(\vec\phi=\vec\phi_{\text{UV}})=\sum_{\bar I}\Lambda_{\bar I}\alpha_{\bar Ii}e^{\vec\alpha_{\bar I}\cdot\vec\phi_{\text{UV}}}+\sum_a\sum_{\tilde I_a}\Lambda_{\tilde I_a}\alpha_{\tilde I_a i}e^{\vec\alpha_{\tilde I_a}\cdot\vec\phi_{\text{UV}}}\,.
\end{equation}

We have thus defined what the values of the potential are in the expressions for the parameters supporting the $\theta=1$ IR solution \eqref{eq:alphax^2}--\eqref{eq:gammaIR}
and in the parameters supporting the $\theta=0$ UV solution \eqref{eq:AdSUV1} and \eqref{eq:AdSUV2} or \eqref{eq:m^2UV}--\eqref{eq:conditionfirstderV_UV}. This provides us with a large class of models in which we can choose $z_{\text{IR}}=z$ and $z_{\text{UV}}=Z$ and look for space-times that interpolate between these two solutions. We note that what we call the IR is a singular solution because of divergent tidal forces \cite{Shaghoulian:2011aa}  and/or because of a divergent dilaton (for example $\theta=1$ and $z=3/2$ has no divergent tidal forces but the dilaton still blows up). Hence we need to further deform the IR to a third geometry (in the spirit of \cite{Harrison:2012vy,Bhattacharya:2012zu,Kundu:2012jn}) which is again a solution of our model \eqref{eq:Lifmodelmulti} that is free of any singularities. We leave the construction of such interpolating solutions to future work.

%%%%%%%%%%%%%%%%%%%%%%%%%%%%%%%%%%%
%%%%%%%%%%%%%%%%%%%%%%%%%%%%%%%%%%%

\section{A Schr\"odinger perspective}
In the previous three sections we took our inspiration from toy model Lagragians that are known for Lifshitz space-times and generalized them appropriately to account for hvLif space-times. In this section, we will briefly discuss a slightly different approach that is also based on massive vector fields, now via dimensional reduction of a Schr\"odinger space-time. We know that a 4D $z=2$ Lifshitz space-time can be obtained by dimensional reduction of a 5D $z=0$ Schr\"odinger space-time \cite{Balasubramanian:2010uk,Donos:2010tu,Cassani:2011sv,Chemissany:2011mb,Chemissany:2012du}. This can be generalized to other values of $z$ except that now we obtain a hvLif space-time from a 5D Schr\"odinger space-time as we will see below.

For Schr\"odinger space-times it is known that a `natural theory' is one containing a massive vector field satisfying all the Schr\"odinger symmetries. Schr\"odinger space-times satisfying the NEC come in two different forms. These are
\begin{eqnarray}
d\hat s^2 & = & \frac{1}{r^{2\zeta}}du^2+2\frac{dudt}{r^2}+\frac{1}{r^2}\left(dr^2+dx^2+dy^2\right)\,,\qquad\text{for}\qquad\zeta^2\le 1\,,\label{eq:Schmetric1}\\
d\hat s^2 & = & -\frac{1}{r^{2\zeta}}du^2+2\frac{dudt}{r^2}+\frac{1}{r^2}\left(dr^2+dx^2+dy^2\right)\,,\qquad\text{for}\qquad\zeta^2\ge 1\,.\label{eq:Schmetric2}
\end{eqnarray}
The class with $\zeta^2\le 1$ can be dimensionally reduced along $u$ by writing 
\begin{equation}\label{eq:KKansatz}
d\hat s^2=e^{2\Phi}\left(du+A^0\right)^2+e^{-\Phi}ds^2\,,
\end{equation}
with 
\begin{eqnarray}
e^{\Phi} & = & r^{-\zeta}\,,\\
A^0 & = & \frac{dt}{r^{2-\zeta}}\,,\\
ds^2 & = & r^{-\zeta}\left(-\frac{dt^2}{r^{2(2-\zeta)}}+\frac{1}{r^2}\left(dr^2+dx^2+dy^2\right)\right)\,.\label{eq:reducedSchmetric1}
\end{eqnarray}
This corresponds to a hvLif space-time with $\theta=-\zeta$ and $z=2-\zeta$. Hence we have $z=2+\theta$. The space-times \eqref{eq:Schmetric1} with $0\le\zeta\le 1$  and $\zeta=-1$ (and not for  $-1<\zeta<0$) are solutions of the following theory
\begin{equation}\label{eq:5Dmassivevectormodel}
S=\int d^5x\sqrt{-\hat g}\left(\hat R-\frac{1}{4}\hat F^2-\frac{1}{2}m^2\hat A^2-V_0\right)\,,
\end{equation}
with 
\begin{eqnarray}
V_0 & = & -12\,,\\
m^2 & = & \zeta(\zeta+2)\,,\\
\hat A & = & \pm\left(\frac{1-\zeta}{2\zeta}\right)^{1/2}\frac{du}{r^{\zeta}}\,,\qquad\text{with}\qquad 0<\zeta\le 1\,.
\end{eqnarray}
The cases $\zeta=0$ and $\zeta=-1$ are special. The latter is a solution of pure gravity with a cosmological constant $V_0=-12$. The former requires to first perform the consistent truncation $\hat A_{\hat\mu}=\frac{1}{m}\partial_{\hat\mu}\hat\varphi$ in the action and the equations of motion with $m$ nonzero. If we reduce this theory down to 4-dimensions with the metric as in \eqref{eq:KKansatz} and with the vector decomposed as
\begin{equation}
\hat A = \chi\left(du+A^0_\mu dx^\mu\right)+A_\mu dx^\mu\,,
\end{equation}
we find
\begin{eqnarray}
S & = & \int d^4x\sqrt{-g}\left(R-\frac{3}{2}\left(\partial\Phi\right)^2-\frac{1}{2}e^{-2\Phi}\left(\partial\chi\right)^2-\frac{1}{4}e^{3\Phi}F^0_{\mu\nu}F^{0\mu\nu}\right.\nonumber\\
&&\left.-\frac{1}{4}e^\Phi\left(F_{\mu\nu}+\chi F^0_{\mu\nu}\right)\left(F^{\mu\nu}+\chi F^{0\mu\nu}\right)-\frac{1}{2}m^2A^2-V_0e^{-\Phi}-\frac{1}{2}m^2\chi^2e^{-3\Phi}\right)\,.\label{eq:4Dactionaxions}
\end{eqnarray}
The action \eqref{eq:4Dactionaxions} provides a different class of models supporting hvLif solutions where an axionic scalar plays an important role. It would be interesting to further work out the potential uses of such a model.

A possible generalization of this method is to start with scale-covariant Schr\"odinger space-times, Ref.~\cite{Kim:2012nb,Perlmutter:2012he}, by adding a dilaton to the 5-dimensional action \eqref{eq:5Dmassivevectormodel} and to reduce in the same way to four dimensions to obtain a larger class of hvLif space-times.

%%%%%%%%%%%%%%%%%%%%%%%%%%%%%%%%%%
%%%%%%%%%%%%%%%%%%%%%%%%%%%%%%%%%%

\section{A waves on branes perspective}\label{sec:wavesonbranes}

In this section, we will explore a different approach to constructing hvLif space-times. We will consider a gravitational wave propagating on the world-volume of an extremal black brane, and take the near-horizon limit. The hvLif metric will result from dimensional reduction%
\footnote{See also e.g. \cite{Gouteraux:2011ce,Gouteraux:2011qh} for similar approaches to hvLif space-times via dimensional reduction.}
on the directions transversal to the brane and on the direction of propagation of the wave along the brane. This procedure will allow for a certain range of values of the scaling exponents $\theta$ and $z$.

We will consider three different scenarios:
\begin{enumerate}
	\item Single $q$-brane with a purely gravitational wave. The brane will act as a source for a $(q+2)$-form field strength.
	
	\item Two intersecting branes with a gravitational wave in the intersection. Each brane will source a different field strength.
	
	\item Single brane with two gauge fields - one which charges the brane and one that  sources the wave.
\end{enumerate}
The main result of the first two cases is given in equations~\eqref{eq:finalintersectingbranes} to \eqref{eq:finalintersectingbranestheta}. The third case is briefly discussed in Subsection~\ref{subsec:wavesourcedbyfluxes}.

We take the branes to have codimensionality 3 at least. The first and the third scenarios are inspired by the work of \cite{Chemissany:2011mb} where gravitational as well as axion waves along extremal D3 branes were considered. 

\subsection{Model and equations of motion}

We consider the case of intersecting branes each acting as an electric source for a $(q_I+2)$-form field strength $F_{q_I+2}$. We thus consider solutions to the following action,
\begin{equation} \label{eqn::braneaction}
	S = \int d^{D}x \sqrt{-g} \left[ R - \frac{1}{2}(\partial \phi)^2 - \sum_I \frac{1}{2(q_I+2)!} e^{a_I \phi} F_{q_I+2}^2 \right].
\end{equation}
We note that there is an ambiguity in the notation for the form fields that occurs when two form fields have the same rank. One should think of the form fields in \eqref{eqn::braneaction} as carrying an additional $I$ label that we have suppressed. The equations of motion are
\begin{eqnarray}
	R_{\mu\nu} &=& \frac{1}{2}\nabla_{\mu} \phi \nabla_{\nu} \phi +  \sum_I \frac{1}{2(q_I+1)!} e^{a_I\phi} F_{\mu\rho_1 \cdots \rho_{q_I+1}}F_{\nu}^{\phantom{\nu}\rho_1 \cdots \rho_{q_I+1}}\nonumber \\
	&&  - \sum_I \frac{q_I+1}{2(D-2)(q_I+2)!} e^{a_I\phi} F_{q_I+2}^2\, g_{\mu\nu} \,, \\
	\nabla_{\mu}\nabla^{\mu} \phi &=& \sum_I \frac{a_I}{2(q_I+2)! }e^{a_I\phi} F_{q_I+2}^2 \,, \\
	0 &=& \partial_{\mu}\left( \sqrt{-g} e^{a_I\phi} F^{\mu\rho_1 \cdots \rho_{q_I+1}} \right)\,.
\end{eqnarray}
We know a large class of extremal solutions (e.g. see \cite{Argurio:1998cp}) for which the metric has the form
\begin{equation}
	ds^2 = - \prod_{I} H_{I}^{-2\frac{D-q_I-3}{\Delta_I}} dt^2 + \sum_{i} \prod_{I} H_{I}^{-2\frac{\delta^i_I}{\Delta_I}} \delta_{ij}dy^idy^j + \prod_{I} H_{I}^{2\frac{q_I+1}{\Delta_I}} \delta_{ab} d\xi^a d\xi^b \,,\label{eq:extremalintersectingbranesmetric}
\end{equation}
where $\delta^i_I$ and $\Delta_I$ are given below. Here $I$ runs over the number of branes in the configuration and the $q_I$ denote the spatial dimensionality of the branes. The brane directions are parametrized by the $y^i$ with $i=1,\ldots,p$. The transverse space is parametrized by the $\xi^a$ coordinates with $a=1,\ldots,n+2$. The total space-time dimension is $D = p + n + 3$.  To each brane is associated a gauge field and a dilaton coupling $a_I$. For an electric ansatz one has the following field strengths,
\begin{equation}
	F_{t i_1 \cdots i_{q_I} a} = \varepsilon_{i_1 \cdots i_{q_I}} \sqrt{\frac{2(D-2)}{\Delta_I}} \partial_a H_I^{-1} \,,
\end{equation}
where $\varepsilon_{i_1 \cdots i_{q_I}}$ is the volume-form of the spatial part of the brane's world-volume. Further, the dilaton is given by
\begin{equation}
	e^{\phi} = \prod_I H_I^{a_I \frac{D-2}{\Delta_I}} \,.
\end{equation}
The harmonic functions can be chosen to describe multi-center solutions, however here we restrict to single brane solutions, in which case we have
\begin{equation}\label{eq:harmonic}
	H_I = 1 +  \frac{r_{0I}}{r^{n}} \,,
\end{equation}
where $r^2=\sum_a(\xi^a)^2$. Furthermore, one defines the following quantities
\begin{equation}\label{eq:DeltaA}
	\Delta_I = (q_I+1)(D-q_I-3) + \frac{1}{2} a_I^2 (D-2) \,,
\end{equation}
and
\begin{equation}
	\delta^i_I = \left\{
	\begin{array}{cl}
		D-q_I-3 & \text{for } i \parallel I \\
		-(q_I + 1) & \text{for } i \perp I
	\end{array} \right. \,,
\end{equation}
where $i \parallel I$ refers to the directions parallel to brane $I$ and $i \perp I$ to the directions perpendicular to brane $I$. The consistency condition for brane intersections is that the dimension of the intersection for each pair of branes $IJ$ is
\begin{equation}
		\bar{q}+1 = \frac{(q_I+1)(q_J+1)}{D-2} - \frac{1}{2} a_I a_J \,,
\end{equation}
where $a_I$ is the dilaton coupling of the corresponding gauge field and $\bar{q}$ is the dimensionality of the pairwise intersection. Notice that these constraint equations relate the dilaton couplings for each pair of intersecting branes. This imposes a strong restriction on the possible configurations of branes.

\subsection{Gravitational wave on a single extremal brane}
\label{sec:SingleBrane}

The theory given by equation (\ref{eqn::braneaction}) admits a class of black $p$-brane solutions carrying $p$-brane charge. These are generalizations of the Gibbons-Maeda black holes \cite{Gibbons:1987ps, Caldarelli:2010xz} and have a single gauge field in the theory that charges the $p$-brane. In this section we will start from the extremal limit of these $D=p+n+3$ dimensional solutions given by equation \eqref{eq:extremalintersectingbranesmetric}. The index $I$ thus only runs over a single value. In this case the metric is
\begin{equation}
	ds^2 = H^{-2\tfrac{D-p-3}{\Delta}} (-dt^2 + d\vec{y}^2 ) + H^{2\tfrac{p+1}{\Delta}}( dr^2 + r^2 d\Omega_{n+1}^2)\,,
	\label{singlebrane}
\end{equation}
where $d\Omega_{n+1}^2$ is the metric on a unit $(n+1)$-sphere. The gauge field $B_{q_I+1}$ defined such that 
\begin{equation}
F_{q_I+2}=dB_{q_I+1} \,,
\end{equation}
with $q_I=p$ is given by
\begin{equation}
	B_{p+1} = \sqrt{\frac{2(D-2)}{\Delta}} \left(\frac{1}{H}-1\right)dt \wedge dy_1 \wedge \cdots \wedge dy_p\,,
\end{equation}
and the dilaton is
\begin{equation}
	\phi = \frac{D-2}{\Delta} a \log H\,,
\end{equation}
where the index $I$ has been dropped on $H$, $\Delta$ and $a$.

We now want to add the gravitational wave for which it is convenient to go to lightcone coordinates defined by
\begin{equation}
	x^{+} = \frac{1}{\sqrt{2}} ( y^1 + t ), \quad x^{-} = \frac{1}{\sqrt{2}} ( y^1 - t )\,.
\end{equation}
Performing the coordinate transformation and adding a gravitational wave (respecting translational symmetries in the $\vec y$-directions), the metric takes the form
\begin{equation}
	ds^2 = H^{-2\tfrac{D-p-3}{\Delta}}\left(2 dx^{+}dx^{-} + d\vec{x}^2 + G(r)(dx^{+})^2 \right) + H^{2\tfrac{p+1}{\Delta}} \left( dr^2 + r^2 d\Omega_{n+1}^2 \right)\,,
\end{equation}
where $\vec{x} = (y^2, ..., y^p)=(x^1, ..., x^{p-1})$. Notice one must have $p\geq1$. The presence of the gravitational wave will neither affect the dilaton equation of motion nor the generalized Maxwell equations and it only shows up in one of the Einstein equations. We have
\begin{equation}
	G(r) = C_1+ \frac{C_2}{r^n}\,. 
\end{equation}
where $C_1$ and $C_2$ are constants. One can choose $C_1=0$ and $C_2=1$ by a coordinate transformation. Let us set $r_0 = 1$ and define
\begin{equation} 
	A=\frac{2n}{\Delta}\,,\qquad B=\frac{2(p+1)}{\Delta}\,,\qquad N = A+B=\frac{2(D-2)}{\Delta} \,.
\end{equation}
We can then take the near horizon limit $r \approx 0$ with the metric written in the form
\begin{equation} 
	ds^2 = r^{2-nB} \left( r^{nN-2}\left[2 dx^{+}dx^{-} + d\vec x^2 + r^{-n}(dx^{+})^2 + r^{-nN} dr^2\right] + d\Omega_{n+1}^2\right)\,. \label{eqn:singlebrane1}
\end{equation}
In order to connect with the canonical form of the hyperscaling violating metrics we define a new radial coordinate $\rho$ via
\begin{eqnarray}
r & = & k\rho^{\tfrac{2}{2-nN}}\,,\qquad k=\left(\frac{2-nN}{2}\right)^{\tfrac{2}{2-nB}}\,,\\
x^+ & = & k^{\tfrac{1-A}{2}n}{x^+}'\,,\\
x^- & = & k^{-\tfrac{1+A}{2}n}{x^-}'\,,\\
\vec x & = & k^{-\tfrac{A}{2}n}{\vec x}'\,.
\end{eqnarray}
This transformation leads to (dropping the prime on the rescaled coordinates)
\begin{eqnarray} 
	\hspace{-1cm}ds^2 & = & \rho^{\frac{Na^2}{2-nN}} \Bigg( \frac{1}{\rho^2} \left[ \rho^{-\frac{2n}{2-nN}} (dx^{+})^2 + d\rho^2 + 2dx^{+}dx^{-} + d\vec x^2 \right] + \left(\frac{2 - nN}{2}\right)^2 d\Omega_{n+1}^2 \Bigg)\,,\label{eqn:GMnh}\\
	\hspace{-1cm}\phi & = & \frac{nNa}{nN-2} \log \rho\,,\\
	\hspace{-1cm}B_{p+1} & = & \sqrt{N}\rho^{\frac{2 n}{ 2 - nN}} dx^{+} \wedge dx^{-} \wedge dx^1 \wedge \cdots \wedge dx^{p-1}\,,
\end{eqnarray}
where we used that the action is invariant under $\phi\rightarrow\phi+c$ and $B_{p+1}\rightarrow e^{-\tfrac{a}{2}c}B_{p+1}$ with $c=\tfrac{1}{2}nNa\log k$.

The conformal factor in equation (\ref{eqn:GMnh}) can be removed by defining a new frame, denoted the dual frame, such that
\begin{equation}
	ds^2 = e^{-\frac{a}{n}\phi} ds^2_{\text{DF}}\,.
\end{equation}
The near horizon metric in the dual frame can be written as
\begin{equation}
	ds^2_{\text{DF}} = \frac{1}{\rho^2} \left( \rho^{-\frac{2n}{2-nN}} (dx^{+})^2 + d\rho^2 - 2dx^{+}dx^{-} + d\vec x^2 \right) + \left(\frac{2 - nN}{2}\right)^2 d\Omega_{n+1}^2\,.
\end{equation}
The action in the dual frame takes the form
\begin{eqnarray}
	S &=& \int d^{D}x \sqrt{-g_{\text{DF}}} \left[ e^{-(\frac{D}{2}-1)\frac{a}{n}\phi} \left( R_{\text{DF}} + \left( - \frac{1}{2} + (D-2)(D-1)\frac{a^2}{4n^2} \right) (\partial \phi)^2  
	\right) \right. \nonumber\\ 
	&& \left. - \frac{1}{2(p+2)!} e^{(\frac{D}{2}-1)\frac{a}{n}\phi} F_{p+2}^2 \right]\,,
\end{eqnarray}
where $R_{\text{DF}}$ is the Ricci scalar of the dual frame metric and where the implicit inverse metric is $g_{\text{DF}}^{\mu\nu}$.

We dualize the $(p+2)$-form field strength to a $(n+1)$-form. The brane is now magnetically charged with respect to the dual field strength which is proportional to the volume form of the transverse $(n+1)$-dimensional sphere. Reducing over the sphere of fixed radius (Freund--Rubin compactification) gives
\begin{eqnarray}
	S &\propto& \int d^{p+2}x \sqrt{-g_{\text{L}}} \left[ e^{-(\frac{D}{2}-1)\frac{a}{n}\phi} \left( R_{\text{L}} + \left( - \frac{1}{2} + (D-2)(D-1)\frac{a^2}{4n^2} \right) (\partial \phi)^2
	\right) \right.\nonumber \\ 
	&& \left. + \frac{4n^2+2n(2-Nn)}{(2-nN)^2} e^{-(\frac{D}{2}-1)\frac{a}{n}\phi}  \right]\,,
\end{eqnarray}
where now the contractions are with respect to the $(p+2)$-dimensional metric $g_L$. The proportionality factor is basically the volume of the transverse sphere over which we reduced the theory. We will continue omitting such volume factors in the reductions below. The metric is given by
\begin{equation} \label{eqn:dualframe}
	ds^2_{\text{L}} = \frac{1}{\rho^2} \left( \rho^{-\frac{2n}{2-nN}} (dx^{+})^2 + d\rho^2 + 2dx^{+}dx^{-} + d\vec x^2 \right)\,.
\end{equation}

One can now go to the Einstein frame by performing a redefinition of the metric of the $(p+2)$-dimensional theory given by equation (\ref{eqn:dualframe}),
\begin{equation}
	ds^2_{\text{L}} = e^{\frac{(D-2)a}{n p}\phi} d\hat s^2_{\text{E}}\,.
\end{equation}
The corresponding action takes the form
\begin{eqnarray}
	S &\propto &\int d^{p+2}x \sqrt{-\hat g_{\text{E}}} \left[\hat R_{\text{E}} -\frac{(D-2)}{2 p n^2N}(2-nN+2n)(\partial \phi)^2 \right.\nonumber\\
	&&\left.+ \frac{4n^2+2n(2-nN)}{(2-nN)^2} e^{\frac{(D-2)a}{np}\phi}   \right]\,,
\end{eqnarray}
and the metric is given by
\begin{equation}
	d\hat s^2_{\text{E}} = \rho^{\frac{(D-2)Na^2}{p(2-nN)}-2}\left( \rho^{-\frac{2n}{2-nN}} (dx^{+})^2 + d\rho^2 + 2dx^{+}dx^{-} + d\vec x^2 \right)\,.
\end{equation}
This is conformal to a Schr\"odinger metric. Note that
\begin{equation}
\frac{(D-2)Na^2}{2(2-nN)}=p+1+\frac{2n}{2-nN}\,.
\end{equation}

Now, we are in position to perform a Kaluza-Klein reduction to go to a ($p+1$)-dimensional theory and obtain the sought-after hyperscaling violating Lifshitz space-times. The reduction is done along the $x^{+}$ direction. The Kaluza-Klein decomposition ansatz in the Einstein frame is
\begin{equation}
	d\hat s^2_{\text{E}}= e^{\frac{1}{(p-1)p} \varphi} ds^2_{\text{E}} + e^{ -\frac{1}{p}\varphi }( dx^{+} + A_{\mu} dx^{\mu} )^2 \,.
\end{equation}
Note that one must have $p\geq2$ and thus the Kaluza-Klein field strength can never be space filling.

The action can be written as
\begin{eqnarray}
	S &\propto& \int d^{p+1}x \sqrt{-g_{\text{E}}} \left[ R - \frac{1}{4(p-1)p}(\partial\varphi)^2  -\frac{(D-2)}{2pn^2N}(2-nN+2n) (\partial \phi)^2   \right. \nonumber  \\
	&& \left. - \frac{1}{4}e^{-\frac{1}{(p-1)}\varphi}F^2 +\frac{4n^2+2n(2-nN)}{(2-nN)^2} e^{\frac{(D-2)a}{np}\phi+\frac{1}{(p-1)p}\varphi} \right]\,.
\end{eqnarray}
We end up with a theory containing four dynamical fields, $g_{\mu\nu}$, $\phi$, $\varphi$ and $A_{\mu}$. The action has two kinetic dilaton terms, two Liouville potentials and a Maxwell gauge field.

The Kaluza-Klein scalar is
\begin{equation}
	\varphi =  \left(-2+2n\frac{p-2}{2-nN}\right)\log\rho\,,
\end{equation}
while the metric is
\begin{equation}
	ds^2_{\text{E}} = \rho^{\frac{2}{p-1}\frac{2+n-nN}{2-nN}} \left( - \rho^{\frac{2n}{2-nN}} (dx^{-})^2 + d\rho^2 + d\vec x^2 \right)\,,
\end{equation}
and the KK vector is
\begin{equation}
	\mathcal{A} = \rho^{\frac{2n}{2-nN}}dx^-\,.
\end{equation}

We want to compare the metric to the space-time with Lifshitz scaling and hyperscaling violation parameters of the form
\begin{equation}
\label{dsztheta}
	ds^2 = \rho^{-2\left(1-\frac{\theta}{p-1}\right)} \left( - \rho^{-2(z-1)} (dx^{-})^2 + d\rho^2 + d\vec x^2 \right).
\end{equation}
One therefore finds the following values for $z$ and $\theta$:
\begin{eqnarray}
	z &=&1+ \frac{n}{n N - 2}\,, \\
	\theta &=& p-\frac{n}{nN-2}\,.  \label{rel1}
\end{eqnarray}
These values are related by
\begin{equation}
\label{ztheta}
z+\theta = p+1 \,,
\end{equation}
where $p+1$ is the dimensionality of the final space-time \eqref{dsztheta}. We note that the sum $z+\theta$ is independent of $n$, so as far as this relation is concerned the `memory' of the dimensionality $n+p+3$ of the initial space-time is lost.

An important question is whether the null energy condition is respected in the space-time \eqref{dsztheta}, which is required if the dual field theory is physically sensible. The null energy condition demands that \cite{Dong:2012se}
\begin{eqnarray}
(d - \theta) (d(z - 1) - \theta) \geq 0 \,, \\
(z - 1)(d + z - \theta) \geq 0 \,,
\end{eqnarray}
where $d=p-1$. Using our expression \eqref{ztheta}, we can see that these inequalities correspond to
\begin{eqnarray}
p(z-2)^2 \geq 0, \\
2(z - 1)^2 \geq 0,
\end{eqnarray}
respectively, and therefore always hold. This is not surprising since the null energy condition is satisfied in the higher-dimensional spacetime, before the reduction. The relation \eqref{ztheta} tells us that the pure Lifshitz case ($\theta=0$) is obtained when $z=p+1$. Our procedure requires $p\geq2$, and thus $z\geq 3$. In particular, AdS space-time ($z=1$) is not allowed.

\subsection{Gravitational wave on two intersecting branes}

In this subsection, we consider the case of intersecting branes each acting as an electric source for a $(q_I+2)$-form field strength $F_{q_I+2}$. We want to relate the brane intersecting configurations to hyperscaling violating metrics by performing successive reductions on them. The final configuration will consist of the time direction and the directions for which all the branes intersect. For simplicity we restrict the analysis to two intersecting branes, so that the label $I$ only assumes the two values 1 and 2. The analysis is straightforward to generalize to multiple intersecting branes that do not have any additional intersections among the branes. The two electrically charged branes have $q_1$ and $q_2$ spatial directions, respectively. The dimensionality of the intersection is denoted $\bar{q} = q_1 \cap q_2$. For two intersecting branes the total number of brane directions is,
\begin{equation}
	p = q_1 + q_2 - \bar{q} \,.
\end{equation}
The metric will then take the form,
\begin{eqnarray}
	ds^2 &=& H_1^{-\alpha_1} H_2^{-\alpha_2} ( -dt^2 + d\vec{y}_{[\bar{q}]}^2 ) +\nonumber \\
	&& +H_1^{-\alpha_1} H_2^{\beta_2} ( d\vec{y}_{[1]}^2 ) +H_1^{\beta_1} H_2^{-\alpha_2} ( d\vec{y}_{[2]}^2 ) \nonumber\\
	&& +H_1^{\beta_1} H_2^{\beta_2} ( dr^2 + r^2 d\Omega^2_{n+1} ),
\end{eqnarray}
where $\vec{y}_{[\bar{q}]}$ denote the $\bar{q}$-directions where the two branes intersect, and $\vec{y}_{[1]}, \vec{y}_{[2]}$ are the $(q_1-\bar{q})$ and $(q_2-\bar{q})$ directions of the branes, respectively. The coefficients are given by,
\begin{eqnarray}
	\alpha_1 = 2 \frac{(D - q_1 - 3)}{\Delta_1}, &\quad& 
	\alpha_2 = 2 \frac{(D - q_2 - 3)}{\Delta_2}, \\
	\beta_1 = 2 \frac{(q_1 + 1)}{\Delta_1}, &\quad&
	\beta_2 = 2 \frac{(q_2 + 1)}{\Delta_2}.
\end{eqnarray}
In order to ensure having a null direction after adding a wave, one must at least have an intersection of dimension $\bar{q} \geq 1$. One can add a gravitational wave in the same way as in Subsection \ref{sec:SingleBrane}, leading to
\begin{eqnarray}
	ds^2 &=& H_1^{-\alpha_1} H_2^{-\alpha_2} ( 2dx^{+}dx^{-} + G(r) (dx^{+})^2 + d\vec{x}^2 )\nonumber  \\
	&&+ H_1^{-\alpha_1} H_2^{\beta_2} ( d\vec{y}_{[1]}^2 ) +  H_1^{\beta_1} H_2^{-\alpha_2} ( d\vec{y}_{[2]}^2 )\nonumber  \\
	&& +H_1^{\beta_1} H_2^{\beta_2} ( dr^2 + r^2 d\Omega^2_{n+1} ),
\end{eqnarray}
In the following, we take $r_{0I} = 1$ (see equation \eqref{eq:harmonic}), such that $H_I = r^{-n}$ for all $I$, and $G(r) = r^{-n}$. Then the near horizon $r \approx 0$ limit of the intersecting branes can be written as
\begin{eqnarray}
	ds^2 &=& r^{2-nB} \Big( r^{nN-2} \left( 2dx^{+}dx^{-} + r^{-n} (dx^{+})^2 + d\vec{x}^2 + r^{-nN} dr^2 \right) \nonumber  \\
	 && +\sum_I r^{n\gamma_I - 2} ( d\vec{y}_{[I]} )^2 + 
	  d\Omega^2_{n+1} \Big),
\end{eqnarray}
where we have defined
\begin{eqnarray}
	A = \sum_{I} \alpha_I, \quad B = \sum_{I} \beta_I, \quad N = \sum_I \gamma_I, \quad \gamma_I = \alpha_I+\beta_I \,.
\end{eqnarray}
This overlaps with the case of a single brane given by equation (\ref{eqn:singlebrane1}), except that now we have additional orthogonal brane directions. In order to connect with the canonical form of the hyperscaling violating metrics, we define, similarly to the previous subsection, a new radial coordinate $\rho$ via
\begin{eqnarray}
r & = & k\rho^{\tfrac{2}{2-nN}}\,,\qquad k=\left(\frac{2-nN}{2}\right)^{\frac{2}{2-nB}}\,,\\
x^+ & = & k^{\tfrac{1-A}{2}n}{x^+}'\,,\\
x^- & = & k^{-\tfrac{1+A}{2}n}{x^-}'\,,\\
\vec{x} & = & k^{-\tfrac{n}{2} A}{\vec x}'\,,\\
\vec{y}_{[I]} & = & k^{-\tfrac{n}{2}(\gamma_I-B)}\vec{x}_{[I]}' \,.
\end{eqnarray}
With this transformation one finds (dropping the prime on the rescaled coordinates),
\begin{eqnarray} 
	ds^2 & = & \rho^{\frac{2(2-Bn)}{2-nN}} \left( \frac{1}{\rho^2} \left[ \rho^{-\frac{2n}{2-nN}} (dx^{+})^2 + d\rho^2 + 2dx^{+}dx^{-} + d\vec{x}^2 \right]\right.\nonumber\\
	&&\left. \sum_I \rho^{\frac{2(n\gamma_I - 2)}{2-nN}} d\vec{x}_{[I]}^2  + \left(\frac{2 - nN}{2}\right)^2 d\Omega_{n+1}^2 \right)\,, \label{eqn:intsec1}\\
	\phi & = & \alpha \log \rho, \quad \alpha =  \frac{2n(D-2)}{nN-2} \sum_I \frac{a_I}{\Delta_I} \,,\label{eq:defalpha}\\
	F_{x^{+}x^{-} i_2 \cdots i_{q_I} \rho} &=& \varepsilon_{i_2 \cdots i_{q_I}} \sqrt{\frac{2(D-2)}{\Delta_I}}  \left( \frac{2}{2-nN} \right) \rho^{\frac{nN+2(n-1)}{2-nN}} \label{eqn:intsec3} \,,
\end{eqnarray}
where we used that the action \eqref{eqn::braneaction} is invariant under $\phi\rightarrow\phi+c$ and $B_{q_I+1}\rightarrow e^{-a_Ic/2}B_{q_I+1}$ with $c=n(D-2)\sum_I\tfrac{a_I}{\Delta_I}\log k$. In the components of the field strength $F_{x^{+}x^{-} i_2 \cdots i_{q_I} \rho}$ the indices $i_2$, etc. are such that $x^{i_2}=(\vec x,\vec x_{[I]})$.

We now outline the reduction procedure that we have gone through and refer to Appendix \ref{sec:detailsIntBranes} for the details. First, we do a redefinition of the metric in order to remove the conformal factor of the transverse sphere followed by a Freund-Rubin compactification. Second, for each brane $I$, we change the frame to remove the conformal factor in front of the orthogonal $q_I$-brane directions, and compactify on the $(q_I-\bar{q})$-directions. Finally, we go back to the Einstein frame, where the theory is now $(q_I+2)$-dimensional, and perform a Kaluza-Klein reduction to obtain a $(q_I + 1)$-dimensional theory.

It is now possible to read off the dynamical critical exponent $z$ and the hyperscaling violation exponent $\theta$ by relating the metric given in equation (\ref{eqn:intbranefinal}) to the canonical form
\begin{equation}\label{eq:finalintersectingbranes}
	ds^2 = \rho^{-2\left(1-\frac{\theta}{\bar{q}-1}\right) } \left( - \rho^{-2(z-1)} (dx^{-})^2 + d\rho^2 + d \vec{x}^2 \right) \,.
\end{equation}
One finds the same structure appearing as in the single brane case,
\begin{eqnarray}
	z &=& 1+\frac{ n}{n N - 2} \,, \\
	\theta &=& \bar{q} - \frac{n}{nN -2} \,.\label{eq:finalintersectingbranestheta}
\end{eqnarray}
The relation between the exponents is
\begin{equation}
	z + \theta = \bar{q} + 1 \,.
\end{equation}
One could expect that the same structure will appear for multiple intersecting branes that do not have additional intersections among them. It would be interesting to investigate cases with more complicated intersections.

\subsection{Waves sourced by fluxes on an extremal brane}\label{subsec:wavesourcedbyfluxes}

Instead of using the second gauge field to charge a second brane as discussed in the previous section, 
we can also use it as a source for the wave that we add to the single brane.  

We would like to do this in a way that only affects the $x^+ x^+$ component of the Einstein equation and nothing else, just like with the gravitational wave considered above. This puts severe restrictions on the second field strength that we now discuss. 

For definiteness we assume in the following that we have a $p$-brane (with charge sourced by $F_{p+2}$) and the
second field strength is denoted by $F_{q+2}$. Then we require
\begin{equation}\label{eq:extraterm}
 (F_{q+2})^\mu{} _{\mu_2 \cdots \mu_{q+2}}(F_{q+2})^{\nu  \mu_2 \cdots \mu_{q+2}} \propto \delta^\mu_{x^+} \delta^\nu_{x^+} 
 \end{equation}
 so that the corresponding stress tensor only has a non-zero $x^+ x^+$ component acting as an extra
 source in the  $x^+ x^+$ component of the  equations of motion for the metric. 
In particular since $g^{x^+ x^+} = 0$ this means that $F_{q+2}^2 = 0$, so this is a null-flux. 
First we observe that this condition implies that $F_{x^- \mu_2 \cdots \mu_{q+2}} = 0$. We furthermore impose
the symmetries ${\cal{L}}_{K} F_{q+2} = 0 $ for the Killing vectors $K$ corresponding to translations 
$\{ \partial_{x^+} , \partial_{x^-}, \partial_i \}$ ($i=2,\ldots,p)$ along the world volume of the $p$-brane, as well as the rotations $\{ x_i \partial_j - x_j \partial_i \}$  in the $(p-1)$-dimensional subspace of the world-volume transverse
to the light-cone directions $(x^+,x^-)$.  This implies that only the following components can be non-zero
\begin{equation}
\label{Fcond}
F_{x^+ 2 \cdots p \alpha_{p+2} \cdots \alpha_{q+2}}    \quad , \quad 
F_{x^+  \alpha_{2} \cdots \alpha_{q+2}} 
\end{equation}
where the indices labelled with $\alpha$ correspond to angular directions of the internal space. 

We now make use of the equation of motion of the field strength $d (\star e^{-2 \bar a \phi} F_{q+2} ) =0$
and assume that the internal space (which we take to be of dimension $n+1$) is an Einstein space. 
One finds that the field strengths in \eqref{Fcond} should be proportional to the volume form on the corresponding
cycles in the internal space. This means that the internal space ${\cal{M}}^{n+1} $ has an $m$-cycle, and depending on which case we are in, the $m$-cycle should of dimension $q-p+1$ or $q+1$. The $r$-dependence on the right hand side of \eqref{eq:extraterm} will turn out to be just a power law. We leave a detailed analysis of the resulting solutions to future work, but it should be clear that this construction enables new phases and a modification of the relation \eqref{rel1} in parameter space. The construction described above includes and generalizes
the one considered in Ref.~\cite{Narayan:2012hk}.

%%%%%%%%%%%%%%%%%%%%%%%%%%%%%%%%%%
%%%%%%%%%%%%%%%%%%%%%%%%%%%%%%%%%%
\section{Discussion}

In this work, we have analyzed in detail several approaches to the
construction of hyperscaling-violating Lifshitz (hvLif) space-times, which
have attracted a lot of interest in the context of applications of the
holographic correspondence to condensed matter systems.

We had two main goals. One was to obtain explicit models allowing for
hvLif space-times with special scaling exponents, in particular for $\theta=1$
and $z=3/2$ (for dual theories in $d=2$ spatial dimensions), which are
particularly interesting from the point of view of applications to
condensed matter. In Section 2, we considered the EPD model with a dilaton and a
massive vector which allows for such solutions. We also studied linearized
perturbations in that model, and used that to construct the natural probe fields to put on a hvLif space-time. We showed that
these probes satisfy a generalized BF bound. It would be interesting to explore these probes further, e.g. by computing 
their two-point functions.

The other main goal was to generalize the different methods of obtaining
hvLif space-times in the literature. The approaches studied in this paper fall into three categories. The first category is a dilaton extension of the
massive vector model used for Lifshitz space-times, the second category goes via the observation that hvLif space-times can be uplifted to Schr\"odinger 
space-times for well chosen $\theta$ and $z$ parameters (and more generally to scale covariant Schr\"odinger space-times) and finally the last approach is based 
on the observation that the Schr\"odinger picture leads to a waves on branes perspective. Our aim was to present a number of motivated
examples whose study could be the start of a survey of the
``hvLif-landscape" in supergravity-like models. It would of course be very interesting and
important to know exactly how we can find real supergravity embeddings for hvLif space-times for large ranges of $\theta$ and $z$ values.

The model considered in Sections 2 and 3 (and generalized in Section 4) has
hvLif solutions with the interesting property that the violation of
hyperscaling can be chosen to be tied with the running of the dilaton. More explicitly we can construct solutions where the dilaton profile $\phi =\phi_0 + \alpha \log r$ is such that
$\alpha$ goes to zero as $\theta$ goes to zero (and similarly for the non-invariance of the 1-form $A$ under Lifshitz rescalings). This requirement selects the EPD model as the preferred model as opposed to the EMD model where this does not hold. The advantage of the EMD model is however that it is possible to construct analytic black hole solutions whereas this may prove to be hard for the EPD model as this has been proven difficult for Lifshitz black hole solutions of the massive vector model. It would be very advantageous to get better analytic control over black brane/hole solutions of theories with massive vector fields as these are needed for many holographic applications\footnote{See e.g. Refs.~
\cite{Danielsson:2009gi,Bertoldi:2009vn,Balasubramanian:2009rx,Tarrio:2011de} for Lifshitz black holes,
\cite{Sadeghi:2012vv,Alishahiha:2012qu,Donos:2012yu} for hyperscaling violating Lifshitz black holes
and \cite{Charmousis:2010zz,Huijse:2011ef,Barisch:2011ui,BarischDick:2012gj} for other interesting black brane solutions in gauged supergravity.} along with their embedding into (gauged) supergravity. Another attractive feature of the EPD model in contrast to the EMD model is that it can account for all values of $\theta$ and $z$ including the case $\theta=1$
and $z=3/2$, which is not a solution of the EMD model. 

We recall that hvLif space-times have pathological IR and
UV limits for generic scaling exponents $\theta$ and $z$, as
discussed in the Introduction. The special case with exponents $\theta=1$
and $z=3/2$ possesses both pathologies: the curvature invariants diverge
in the UV (since $\theta>0$); in the IR, while the metric is regular, the
dilaton runs logarithmically. As stressed in \cite{Dong:2012se}, the hvLif
space-time should be an effective description for intermediate scales, not
too close to the IR or to the UV. One interesting extension of our work
would be to construct an explicit interpolating solution, where our hvLif
solutions are provided with a regular IR geometry (such as $AdS_2 \times R^2$) and
a regular UV geometry which can be used for holographic computations
(such as $AdS_4$ or a Lifshitz space-time). We have already studied in
Section 4 a model that should at least allow for a UV completion. However,
we point out that, in principle, one should be able to use
holography without such explicit completions, in the spirit of effective
field theories. This extension of the holographic dictionary would be
crucial not only for practical applications, but also for developing a deeper
understanding of the holographic correspondence.

%%%%%%%%%%%%%%%%%%%%%%%%%%%%%%%%%%
\section*{Acknowledgements}

We thank Wissam Chemissany and Blaise Gout\'eraux for useful discussions. 
The work of JH and NO is supported in part by the Danish National Research Foundation project ``Black holes and their role in quantum gravity''.

%%%%%%%%%%%%%%%%%%%%%%%%%%%%%%%%%%
%%%%%%%%%%%%%%%%%%%%%%%%%%%%%%%%%%

\appendix
\section{Linearized equations}

In this appendix, we present lists of equations used in Section \ref{sec:linear}, when analyzing perturbations of the hvLif background \eqref{eq:phim^2neq0}--\eqref{eq:am^2neq0} for general $z$ and $\theta$.

\subsection{List of equations of motion}
\label{subsec1:appendix}

We list here the linearized radial equations of motion in radial gauge, using the parametrization \eqref{eq:metricpertb2}--\eqref{eq:scalarpertb2}, except that we now use $a_\mu$, which is related to $\tilde a_\mu$ through $\tilde a_\mu = \bar A_t r^z a_\mu$. We start with \eqref{eq:firstorderscalar},
\begin{eqnarray}
0 & = & -r^{2z}\partial_t^2\varphi+r^2\partial_r^2\varphi+(\theta-z-1)r\partial_r\varphi+r^2(\partial_x^2\varphi+\partial_y^2\varphi)+\frac{x}{2}r\partial_r h\nonumber\\
&&-\left[(z-2)(z-1)ab-2(z-1)za^2-\theta(\theta-z-2)\right]\varphi\nonumber\\
&&+\left[\frac{1}{2}(z-1)(\theta-4)b-(z-1)za\right]h^t{}_t-2a(z-1)r^{z+1}(\partial_r a_t-\partial_t a_r)\nonumber\\
&&-b(z-1)(\theta-4)r^za_t\,.
\end{eqnarray}
The $t$ component of \eqref{eq:firstordervector} gives
\begin{eqnarray}
0 & = & \kappa a r\partial_r\varphi+(b-a)\kappa\left(2-\theta-\kappa+z\right)\varphi-r^{z+2}\partial_r^2a_t+r^{z+2}\partial_r\partial_ta_r\nonumber\\
&&+\left(\theta-1+\kappa\right)r^{z+1}\partial_t a_r-(z-1+\theta)r^{z+1}\partial_ra_t-z(\theta-2)r^za_t\nonumber\\
&&-r^{z+2}\partial_xf_{xt}-r^{z+2}\partial_y f_{yt}-\kappa r\partial_r h^t{}_t+\frac{1}{2}\kappa r\partial_r h\,.
\end{eqnarray}
For the $r$ component of \eqref{eq:firstordervector} we find
\begin{eqnarray}
0 & = & -\kappa ar^z\partial_t\varphi-r^{2z+1}\partial_t^2a_r+r^3(\partial_x^2a_r+\partial_y^2a_r)-\kappa\left(2-\theta-\kappa+z\right)ra_r\nonumber\\
&&+r^{2z+1}\partial_t\partial_r a_t+(z-\kappa)r^{2z}\partial_t a_t-r^3\partial_x\partial_r a_x-r^3\partial_y\partial_r a_y+\kappa r^z\partial_a h^a{}_t-\frac{\kappa}{2}r^z\partial_t h\,.
\end{eqnarray}
For the $x$ component of \eqref{eq:firstordervector} we find
\begin{eqnarray}
0 & = & r^3(\partial_r^2a_x-\partial_r\partial_xa_r)+(\theta-z+1)r^2\partial_r a_x-\left(\theta-2z+1+\kappa\right)r^2\partial_x a_r\nonumber\\
&&+\left(bx(z-1)+z(\theta-2)\right)ra_x-r^{2z+1}\partial_t f_{tx}-r^3\partial_y f_{xy}-\kappa r^z\partial_r h^x{}_t\,.
\end{eqnarray}
For the $y$ component of \eqref{eq:firstordervector} we find
\begin{eqnarray}
0 & = & r^3(\partial_r^2a_y-\partial_r\partial_ya_r)+(\theta-z+1)r^2\partial_r a_y-\left(\theta-2z+1+\kappa\right)r^2\partial_y a_r\nonumber\\
&&+\left[2(z-1)\kappa+z(\theta-2z)\right]ra_y-r^{2z+1}\partial_t f_{ty}+r^3\partial_x f_{xy}-\kappa r^z\partial_r h^y{}_t\,.
\end{eqnarray}
The $tt$ component of the Einstein equations is
\begin{eqnarray}
0 & = & (z-1)(2\theta-2z-4+\kappa)h^t{}_t-\frac{1}{2}(3\theta-4z-2)r\partial_rh^t{}_t-\frac{1}{2}(\theta-2z)r\partial_r\left(h^x{}_x+h^y{}_y\right)\nonumber\\
&&-r^2\left(\partial_y^2h^t{}_t+\partial_x^2h^t{}_t+\partial_r^2h^t{}_t\right)-2r^{2z}\left(\partial_t\partial_y h^y{}_t+\partial_t\partial_x h^x{}_t\right)+r^{2z}\partial_t^2\left(h^x{}_x+h^y{}_y\right)\nonumber\\
&&+2(z-1)(4+z-2\theta-\kappa)r^za_t+2(z-1)r^{z+1}(\partial_ta_r-\partial_ra_t)\nonumber\\
&&+\left[(\theta-2-z)\left[b(\theta-2z)-a(\theta-2)\right]-(z-1)b\kappa \right]\varphi\,.
\end{eqnarray}
The $tr$ component of the Einstein equations is
\begin{eqnarray}
0 & = & -2(z-1)r^z\left(\partial_y h^y{}_t+\partial_x h^x{}_t\right)+(z-1)r^z\partial_t\left(h^x{}_x+h^y{}_y\right)-r^{z+1}\left(\partial_r\partial_y h^y{}_t+\partial_r\partial_x h^x{}_t\right)\nonumber\\
&&+r^{z+1}\partial_r\partial_t\left(h^x{}_x+h^y{}_y\right)+xr^z\partial_t\varphi+2(z-1)(2-\theta+z-\kappa)ra_r\,.
\end{eqnarray}
The $tx$ component of the Einstein equations is
\begin{eqnarray}
0 & = & (\theta+z-3)r^z\partial_rh^x{}_t+r^{z+1}\partial_y^2h^x{}_t-r^{z+1}\partial_x\partial_yh^y{}_t+r^{z+1}\partial_r^2h^x{}_t-r^{z+1}\partial_t\partial_yh^x{}_y\nonumber\\
&&+r^{z+1}\partial_t\partial_x h^y{}_y-2(z-1)(\theta-2)ra_x+2(z-1)r^2\left(\partial_xa_r-\partial_ra_x\right)\,.
\end{eqnarray}
The $ty$ component of the Einstein equations is
\begin{eqnarray}
0 & = & (\theta+z-3)r^z\partial_rh^y{}_t+r^{z+1}\partial_x^2h^y{}_t-r^{z+1}\partial_x\partial_yh^x{}_t+r^{z+1}\partial_r^2h^y{}_t-r^{z+1}\partial_t\partial_xh^x{}_y\nonumber\\
&&+r^{z+1}\partial_t\partial_y h^x{}_x-2(z-1)(\theta-2)ra_y+2(z-1)r^2\left(\partial_ya_r-\partial_ra_y\right)\,.
\end{eqnarray}
The $rr$ component of the Einstein equations is
\begin{eqnarray}
0 & = & r^2\partial_r^2h+\frac{1}{2}(\theta-2)r\partial_rh-2(z-1)r\partial_r h^t{}_t+2xr\partial_r\varphi-\frac{\theta}{x}V_0\varphi-\kappa a(z-1)\varphi\nonumber\\
&&-(z-1)bxr^z a_t+2(z-1)r^{z+1}\left(\partial_r a_t-\partial_t a_r\right)+\kappa (z-1)h^t{}_t\,.
\end{eqnarray}
The $rx$ component of the Einstein equations is
\begin{eqnarray}
0 & = & -(z-1)r\partial_xh^t{}_t-r^2\partial_r\partial_yh^x{}_y+r^2\partial_r\partial_xh^t{}_t+r^2\partial_r\partial_xh^y{}_y+r^{2z}\partial_r\partial_th^x{}_t+xr\partial_x\varphi\nonumber\\
&&+2(z-1)r^{z+1}f_{xt}\,.
\end{eqnarray}
The $ry$ component of the Einstein equations is
\begin{eqnarray}
0 & = & -(z-1)r\partial_yh^t{}_t-r^2\partial_r\partial_xh^x{}_y+r^2\partial_r\partial_yh^t{}_t+r^2\partial_r\partial_yh^x{}_x+r^{2z}\partial_r\partial_th^y{}_t+xr\partial_y\varphi\nonumber\\
&&+2(z-1)r^{z+1}f_{yt}\,.
\end{eqnarray}
The $xx$ component of the Einstein equations is
\begin{eqnarray}
0 & = & \frac{1}{2}(\theta-2)r\partial_r h^t{}_t+\frac{1}{2}(3\theta-2z-4)r\partial_rh^x{}_x+\frac{1}{2}(\theta-2)r\partial_rh^y{}_y+r^2\partial_y^2h^x{}_x-2r^2\partial_x\partial_yh^x{}_y\nonumber\\
&&+r^2\partial_x^2h^t{}_t+r^2\partial_x^2h^y{}_y+r^2\partial_r^2h^x{}_x+2r^{2z}\partial_t\partial_xh^x{}_t-r^{2z}\partial_t^2h^x{}_x-(z-1)\kappa h^t{}_t\nonumber\\
&&+\left[(b-a)V_0+(z-1)\kappa a\right]\varphi-2(z-1)(z-\kappa)r^za_t-2(z-1)r^{z+1}\left(\partial_r a_t-\partial_t a_r\right)\,.
\end{eqnarray}
The $xy$ component of the Einstein equations is
\begin{eqnarray}
0 & = & (\theta-z-1)r\partial_rh^x{}_y+r^2\partial_x\partial_yh^t{}_t+r^2\partial_r^2h^x{}_y+r^{2z}\partial_t\partial_yh^x{}_t+r^{2z}\partial_t\partial_xh^y{}_t-r^{2z}\partial_t^2h^x{}_y\,.
\end{eqnarray}
The $yy$ component of the Einstein equations is
\begin{eqnarray}
0 & = & \frac{1}{2}(\theta-2)r\partial_r h^t{}_t+\frac{1}{2}(3\theta-2z-4)r\partial_rh^y{}_y+\frac{1}{2}(\theta-2)r\partial_rh^x{}_x+r^2\partial_x^2h^y{}_y-2r^2\partial_x\partial_yh^x{}_y\nonumber\\
&&+r^2\partial_y^2h^t{}_t+r^2\partial_y^2h^x{}_x+r^2\partial_r^2h^y{}_y+2r^{2z}\partial_t\partial_yh^y{}_t-r^{2z}\partial_t^2h^y{}_y-(z-1)\kappa h^t{}_t\nonumber\\
&&+\left[(b-a)V_0+(z-1)\kappa a\right]\varphi-2(z-1)(z-\kappa)r^za_t-2(z-1)r^{z+1}\left(\partial_r a_t-\partial_t a_r\right)\,.
\end{eqnarray}
There are 15 equations and 11 unknowns. Note that the gauge freedom is still not completely fixed, so some perturbations could still be pure gauge. These equations are all invariant under $r\rightarrow\lambda r$, $t\rightarrow\lambda^z t$, $\vec x\rightarrow\lambda\vec x$ with the fields transforming as
\begin{eqnarray}
h^i{}_j & \rightarrow & h^i{}_j\,,\label{eq:scaling1}\\
h^t{}_t & \rightarrow & h^t{}_t\,,\\
h^t{}_i & \rightarrow & \lambda^{z-1}h^t{}_i\,,\\
h^i{}_t & \rightarrow & \lambda^{-z+1}h^i{}_t\,,\\
a_r & \rightarrow & \lambda^{-1}a_r\,,\\
a_t & \rightarrow & \lambda^{-z}a_t\,,\\
a_i & \rightarrow & \lambda^{-1}a_i\,,\\
\varphi & \rightarrow & \varphi\,.\label{eq:scaling8}
\end{eqnarray}

\subsection{List of matrix components}
\label{subsec2:appendix}

We list here the components of the matrix \eqref{eq:Mscalar}:

\begin{eqnarray}
M_4^{\phantom{s}1} & = & -\frac{1}{2 \alpha^2 (z-1) (-\theta +z+1)} \Big[\alpha^6+\alpha^4 \left(-2 \left(\theta ^2+\theta -3\right)+3 z^2+3 z\right) \nonumber \\
&&+\theta  \alpha^2 \left(\theta ^3+7 \theta ^2-24 \theta +6 z^3-8 (\theta -1) z^2 + \left(\theta ^2-12 \theta +18\right)
   z+16\right) \nonumber\\
&& +2 \theta ^2 (-\theta +z+1)^2 \left(-2 \theta +2 z^2-\theta  z+4\right)\Big] \,,\\
&&\nonumber\\
M_4^{\phantom{s}2} & = & \frac{1}{\alpha (1+z-\theta)} \Big[-\alpha^4+\alpha^2 \left(4 \theta -3 z^2+(2 \theta -3) z-6\right)\nonumber \\
&&+\theta  \left(\theta ^3-7 \theta ^2+14 \theta -2 z^3+4 (\theta -1) z^2+\left(-3 \theta ^2+10 \theta -10\right) z-8\right)  \Big] \,, \\
&&\nonumber\\
M_4^{\phantom{s}3} & = & -\frac{1}{2} M_4^{\phantom{s}2} \,,  \\
&&\nonumber\\
M_4^{\phantom{s}4} & = & \frac{1}{2 (1+z-\theta)} \Big[ \alpha^2+2 \left(\theta ^2-3 \theta +z^2+(3-2 \theta ) z+2\right) \Big] \,, \\
&&\nonumber\\
M_4^{\phantom{s}5} & = & \frac{1}{\alpha (1+z-\theta)} \Big[ \alpha^2 (-2 \theta +3 z+1)+2 \theta  (-\theta +z+1)^2 \Big] \,, \\
&&\nonumber\\
M_4^{\phantom{s}6} & = & \frac{1}{2 (1+z-\theta)} \Big[ \alpha(1-z) \Big] \,,
\end{eqnarray}

\begin{eqnarray}
M_5^{\phantom{s}1} & = & \frac{1}{4 \alpha (z-1)^2 (1+z-\theta)} \Big[\alpha^6+\alpha^4 \left(-4 \theta ^2+5 \theta +3 z^2+(3 \theta -5) z+2\right)\nonumber \\
&& + \alpha^2 \big( \theta  \left(5 \theta ^3-13 \theta ^2+2 \theta +8\right)+2 z^4+8 (\theta -1) z^3+\left(-7 \theta ^2-2 \theta +10\right) z^2 \nonumber\\
&& -\left(7 \theta ^3-25 \theta ^2+14 \theta +4\right)  z\big)+2 \theta  \big(-(\theta -2)^2 \theta  \left(\theta ^2-1\right)+2 z^5-6 z^4\nonumber\\
&&+\left(-5 \theta ^2+12 \theta +2\right) z^3+2 \left(\theta ^3+\theta ^2-8 \theta +3\right) z^2 \nonumber\\
&& +\left(2 \theta
   ^4-11 \theta ^3+15 \theta ^2-4\right) z\big)\Big] \,,  \\
&&\nonumber\\
M_5^{\phantom{s}2} & = & \frac{1}{2 (z-1) (1+z-\theta)} \Big[ (\theta -2)^2 \theta  (\theta +1)+\alpha^4+\alpha^2 \left(-2 \theta ^2+3 \theta +3 z^2+(\theta -5) z+2\right) \nonumber \\
&& +2 z^4+2 (\theta -4) z^3+\left(-3 \theta ^2+2 \theta +10\right) z^2-\left(\theta ^3-7
   \theta ^2+8 \theta +4\right) z \Big] \,, \\
&&\nonumber\\
M_5^{\phantom{s}3} & = & -\frac{1}{2} M_5^{\phantom{s}2} \,,\\
&&\nonumber\\
M_5^{\phantom{s}4} & = & \frac{1}{4 \alpha (z-1)^2 (1+z-\theta)} \Big[ \alpha^4 (-2 \theta +z+3) +2 \theta  (-\theta +z+1)^2   \left(-(\theta -2) \theta +2 z^2-2 z\right) \nonumber \\
&& +\alpha^2 \left(\theta  \left(4 \theta ^2-11 \theta +8\right)+2 z^3-2 (\theta -2) z^2+\left(-5 \theta ^2+10 \theta -6\right) z\right)  \Big] \,,\\
&&\nonumber\\
M_5^{\phantom{s}5} & = & -\frac{1}{2 (1+z-\theta)} \Big[ \alpha^2+(\theta -2) (-3 \theta +4 z+2) \Big] \,,\\
&&\nonumber\\
M_5^{\phantom{s}6} & = & -\frac{1}{4 (z-1) (1+z-\theta)} \Big[ (-2 \theta +z+3) \left(-(\theta -2) \theta +\alpha^2+2 z^2-2 z\right) \Big] \,,\nonumber \\
M_6^{\phantom{s}1} & = & \frac{1}{2\alpha (1+z-\theta)} \Big[ \alpha^4+\alpha^2 \left(-6 \theta +4 z^2-3 (\theta -2) z+8\right) \nonumber \\
&& +2 \theta  \left(2 \left(\theta ^2-3 \theta +2\right)+2 z^3+(2-3 \theta ) z^2+\left(\theta ^2-3 \theta +4\right) z\right)  \Big] \,,\\
&&\nonumber\\
M_6^{\phantom{s}2} & = & \frac{1}{1+z-\theta} \Big[ (z-1) \left(2 (\theta -2)^2+\alpha^2+4 z^2+(6-5 \theta ) z\right) \Big] \,,\\
&&\nonumber\\
M_6^{\phantom{s}3} & = & -\frac{1}{2} M_6^{\phantom{s}2} \,,\\
&&\nonumber\\
M_6^{\phantom{s}4} & = & \frac{1}{2 (1+z-\theta)} \Big[ \alpha (\theta -2 z) \Big] \,,\\
&&\nonumber\\
M_6^{\phantom{s}5} & = & -\frac{1}{2 (1+z-\theta)} \Big[ (z-1) (-3 \theta +4 z+2)\Big] \,,\\
&&\nonumber\\
M_6^{\phantom{s}6} & = & \frac{1}{2 (1+z-\theta)} \Big[ 2 \theta ^2-5 \theta +4 z^2+(4-5 \theta ) z+4\Big] \,.
\end{eqnarray}

\section{Details on intersecting branes}
\label{sec:detailsIntBranes}

In this section we provide the details for the successive reduction procedure performed on the configuration given by equations (\ref{eqn:intsec1})-(\ref{eqn:intsec3}). 

The first step is to go to the dual frame where we will apply the Freund-Rubin compactification on the transverse sphere. The redefinition of the metric is
\begin{equation}
	ds^2 = e^{\omega_0 \phi} ds^2_{\text{DF}} \,,
\end{equation}
for which one has
\begin{equation}
	\omega_0 = \frac{1}{\alpha} \frac{2(2-Bn)}{2-nN} = - \frac{2-nB}{n(D-2)} \left( \sum_I \frac{a_I}{\Delta_I} \right)^{-1} \,,
\end{equation}
where $\alpha$ is defined in equation \eqref{eq:defalpha}. The metric in the dual frame is
\begin{eqnarray} 
	ds^2_{\text{DF}} &=& \frac{1}{\rho^2} \left[ \rho^{-\frac{2n}{2-nN}} (dx^{+})^2 + d\rho^2 + 2dx^{+}dx^{-} + d\vec{x}^2 \right] \nonumber \\
	&& +\sum_I \rho^{\frac{2(n\gamma_I-2)}{2-nN}} d\vec{x}_{[I]}^2  + \left(\frac{2 - nN}{2}\right)^2 d\Omega_{n+1}^2 \,,	
\end{eqnarray}
and the action takes the form
\begin{eqnarray}
	S &=& \int d^{D}x \sqrt{-g_{\text{DF}}} \bigg[ e^{\left(\frac{D}{2}-1\right)\omega_0 \phi} \left( R_{\text{DF}} + \left( - \frac{1}{2} + \frac{(D-2)(D-1)}{4}\omega_0^2 \right) (\partial \phi)^2 \right) \nonumber \\
	&& - \sum_I \frac{1}{2(q_I+2)!} e^{\left( (\frac{D}{2}-(q_I+2))\omega_0 + a_I\right) \phi} F_{q_I+2}^2 \bigg] \,.
\end{eqnarray}
Here $R_{\text{DF}}$ is the Ricci scalar of the dual frame metric and the implicit metric is $g_{\text{DF}}$. We can now reduce over the sphere of fixed radius leading to a theory in $p+2$ dimensions with action
\begin{eqnarray}
	S &=& \int d^{p+2}x \sqrt{-g_{\text{L}}} \bigg[ e^{\left(\frac{D}{2}-1\right)\omega_0\phi} \left( R_{\text{L}} + \left( - \frac{1}{2} + \frac{(D-2)(D-1)}{4}\omega_0^2 \right) (\partial \phi)^2 + \frac{4n(n+1)}{(2-nN)^2} \right) \nonumber \\
	&& - \sum_I \frac{1}{2(q_I+2)!} e^{\left( (\frac{D}{2}-(q_I+2))\omega_0 + a_I\right) \phi} F_{q_I+2}^2 \bigg] \,,
\end{eqnarray}
where the contractions are with respect to the metric $g_{\text{L}}$,
\begin{eqnarray} 
	ds^2_{\text{L}} &=& \frac{1}{\rho^2} \left[ \rho^{-\frac{2n}{2-nN}} (dx^{+})^2 + d\rho^2 + 2dx^{+}dx^{-} + d\vec{x}^2 \right] +
	 \sum_I \rho^{\frac{2(n\gamma_I-2)}{2-nN}} d\vec{x}_{[I]}^2   \,.
\end{eqnarray}

The next step is to perform successive reductions over the orthogonal directions of the branes in the configuration. We define the constants
\begin{eqnarray}
	d_0 &=& p+2 = q_1 + q_2 - \bar{q} +2 \,, \\ 
	d_1 &=& d_0 - q_1 + \bar{q} = q_2 + 2 \,, \\
	d_2 &=& d_1 - q_2 + \bar{q} = \bar{q} + 2 \,,
\end{eqnarray}
which is the dimensionality of the theory at each step of the reduction chain. Also, in the following it is convenient to define the conformal factor of the orthogonal brane directions
\begin{equation} \label{eqn:OmegaI}
	\Omega_I = \frac{2(n\gamma_I - 2)}{2-nN} \,.
\end{equation}
The first frame redefinition is
\begin{equation}
	ds^2_{\text{L}} = e^{\omega_1 \phi} d\hat{s}^2_{[1]}, \quad \omega_{1} = \frac{\Omega_1}{\alpha} \,.
\end{equation}
Here $\Omega_1$ is the conformal factor of the first set of orthogonal brane directions we want to reduce on. In this frame, the metric is
\begin{equation} 
	d\hat{s}_{[1]}^2 = \rho^{-\Omega_1} \left( \frac{1}{\rho^2} \left[ \rho^{-\frac{2n}{2-nN}} (dx^{+})^2 + d\rho^2 + 2dx^{+}dx^{-} + d\vec{x}^2 \right] +
	 \rho^{\Omega_2} d\vec{x}_{[2]}^2 \right) + d\vec{x}_{[1]}^2   \,,
\end{equation}
and the corresponding action is
\begin{eqnarray}
	S &=& \int d^{d_0}x \sqrt{-\hat{g}_{[1]} } \bigg[ e^{s_1 \phi} \left( \hat{R}_{[1]} + s_{\phi} (\partial \phi)^2 \right) + e^{s_{\Omega} \phi} \frac{4n(n+1)}{(2-nN)^2} \nonumber \\
	&& - \sum_I \frac{1}{2(q_I+2)!} e^{s_{F_I} \phi} F_{q_I+2}^2 \bigg] \,,
\end{eqnarray}
where contractions are with the hatted $d_0$-dimensional metric $\hat{g}_1$ and the coefficients are
\begin{eqnarray}
 s_1 &=& \left(\frac{D}{2}-1\right)\omega_0 + \left(\frac{d_0}{2}-1\right)\omega_1 \,, \\ 
  s_{\phi} &=& - \frac{1}{2} + \frac{(D-2)(D-1)}{4}\omega_0^2 + \lambda_1 \,, \\
 s_{\Omega} &=& \left(\frac{D}{2}-1\right)\omega_0 + \frac{d_0}{2}\omega_1 \,, \\
 s_{F_I} &=& a_I + \left(\frac{D}{2}-(q_I+2) \right)\omega_0 + \left(\frac{d_0}{2}-(q_I+2) \right) \omega_1 \,,\\
  \lambda_1 &=& - \frac{(d_0-2)(d_0-1)}{4}\omega_1^2 + (d_0+1)\omega_1 s_1 \,.
\end{eqnarray}
The reduction over the orthogonal brane directions is now trivial and the action takes the form
\begin{eqnarray}
	S &=& \int d^{d_1}x \sqrt{-g_{[1]} } \bigg[ e^{s_1 \phi} \left[ R_{[1]} + s_{\phi} (\partial \phi)^2 \right] + e^{s_{\Omega} \phi} \frac{4n(n+1)}{(2-nN)^2} \nonumber \\
	&& - \frac{1}{2(q_2+2)!} e^{s_{F_2} \phi} F_{q_2+2}^2 - \frac{1}{2(\bar q+2)!}  e^{s_{F_1} \phi} F_{\bar{q}+2}^2 \bigg] \,,
\end{eqnarray}
where the action now has a $(\bar{q}+2)$-form field strength and the metric is
\begin{equation} 
	ds_{[1]}^2 = \rho^{-\Omega_1} \left( \frac{1}{\rho^2} \left[ \rho^{-\frac{2n}{2-nN}} (dx^{+})^2 + d\rho^2 + 2dx^{+}dx^{-} + d\vec{x}^2 \right] +
	  \rho^{\Omega_2} d\vec{x}_{[2]}^2 \right) \,.
\end{equation}
The second frame redefinition will undo the previous one and implement the current one,
\begin{equation}
	ds^2_{[1]} = e^{\omega_2 \phi} d\hat{s}^2_{[2]}, \quad \omega_2 = \frac{ - \Omega_1 + \Omega_2}{\alpha} \,,
\end{equation}
which gives the metric
\begin{equation} 
	d\hat{s}_{[2]}^2 = \rho^{-\Omega_2} \frac{1}{\rho^2} \left[ \rho^{-\frac{2n}{2-nN}} (dx^{+})^2 + d\rho^2 + 2dx^{+}dx^{-} + d\vec{x}^2 \right] + d\vec{x}_{[2]}^2
	   \,.
\end{equation}
The action is now given by 
\begin{eqnarray}
	S &=& \int d^{d_1}x \sqrt{-\hat{g}_{[2]} } \bigg[ e^{s_2 \phi} \left[ \hat{R}_{[2]} + s_{\phi} (\partial \phi)^2 \right] + e^{s_{\Omega} \phi} \frac{4n(n+1)}{(2-nN)^2} \nonumber \\
	&& - \frac{1}{2(\bar q+2)!} e^{s_{F_1} \phi} F_{\bar{q}+2}^2 - \frac{1}{2(q_2+2)!} e^{s_{F_2} \phi} F_{q_2+2}^2 \bigg] \,,
\end{eqnarray}
with the coefficients
\begin{eqnarray}
  s_2 &=& \left(\frac{D}{2}-1\right)\omega_0 + \left(\frac{d_0}{2}-1\right)\omega_1 + \left(\frac{d_1}{2}-1\right)\omega_2 \,, \\
   s_{\phi} &=& - \frac{1}{2} + \frac{(D-2)(D-1)}{4}\omega_0^2 + \lambda_1 + \lambda_2 \,, \\
 s_{\Omega} &=& \left(\frac{D}{2}-1\right)\omega_0 + \frac{d_0}{2}\omega_1 + \frac{d_1}{2}\omega_2 \,, \\
 s_{F_1} &=& a_1 + \left(\frac{D}{2}-(q_1+2) \right)\omega_0 + \left(\frac{d_0}{2}-(q_1+2) \right) \omega_1 + \left(\frac{d_1}{2}-(\bar{q}+2) \right) \omega_2 \,, \\
 s_{F_2} &=& a_2 + \left(\frac{D}{2}-(q_2+2) \right)\omega_0 + \sum_{I=1}^2 \left(\frac{d_{I-1}}{2}-(q_2+2) \right) \omega_I \,,\\
  \lambda_1 &=& - \frac{(d_0-2)(d_0-1)}{4}\omega_1^2 + (d_0+1)\omega_1 s_1  \,, \\
 \lambda_2 &=& - \frac{(d_1-2)(d_1-1)}{4}\omega_2^2 + (d_1+1)\omega_2 s_2  \,, \\
 s_1 &=& \left(\frac{D}{2}-1\right)\omega_0 + \left(\frac{d_0}{2}-1\right)\omega_1 \,.
\end{eqnarray}
Again, the reduction over the second set of orthogonal brane directions is now trivial and the action takes the form
\begin{eqnarray}
	S &=& \int d^{\bar{q}+2}x \sqrt{- g_{[2]} } \bigg[ e^{s_2 \phi} \left[ R_{[2]} + s_{\phi} (\partial \phi)^2 \right] + e^{s_{\Omega} \phi} \frac{4n(n+1)}{(2-nN)^2} \nonumber \\
	&& - \sum_I \frac{1}{2(\bar q+2)!} e^{s_{F_I} \phi} F_{\bar{q}+2}^2 \bigg] \,,\label{eq:intermediateaction}
\end{eqnarray}
where the implicit contractions are with respect to the metric $g_{[2]}$. As mentioned below equation \eqref{eqn::braneaction} we are suppressing an $I$ label on the field strengths. This means that in the sum in \eqref{eq:intermediateaction} each field strength is different. For our solution we have
\begin{equation} 
	ds_{[2]}^2 = \rho^{-\Omega_2} \frac{1}{\rho^2} \left[ \rho^{-\frac{2n}{2-nN}} (dx^{+})^2 + d\rho^2 + 2dx^{+}dx^{-} + d\vec{x}^2 \right] \,.
\end{equation}
This completes the reductions over the orthogonal brane directions and we can now finally go back to Einstein frame using
\begin{equation} \label{eqn:OmegaE}
	ds^2_{[2]} = e^{\omega_{\text{E}} \phi} d\hat{s}^2_{\text{E}}, \quad \omega_{\text{E}} = \frac{\Omega_{\text{E}}}{\alpha} \,.
\end{equation}
with
\begin{eqnarray} 
	\omega_{\text{E}} &=& -\frac{2}{\bar{q}} \left[ \left( \frac{D}{2}-1 \right) \omega_0 + \left( \frac{d_{0}}{2}-1 \right) \omega_1 + \left( \frac{d_{1}}{2}-1 \right) \omega_2 \right] \,, \\
	 &=& -\frac{2}{\bar{q} (2-nN)\alpha} \left[ \left( D-2 \right) (2-Bn) + \left(q_1 - \bar{q} \right) (n\gamma_1 -2) + q_2 (n\gamma_2-2) \right] \,.
\end{eqnarray}
The metric is then given by
\begin{eqnarray}
	d\hat{s}^2_{\text{E}} &=& \rho^{-(\Omega_2 + \Omega_E)} \frac{1}{\rho^2} \left( \rho^{-\frac{2n}{2-nN}} (dx^{+})^2 + 2dx^{+}dx^{-} + d\rho^2 + d \vec{x}^2 \right) \,.
\end{eqnarray}
All the gauge field strengths in \eqref{eq:intermediateaction} are now top-forms. Updualizing them we obtain
\begin{eqnarray}
	S &=& \int d^{\bar{q}+2}x \sqrt{- \hat{g}_{\text{E}} } \bigg[ \hat{R}_{\text{E}} - s_{\phi} (\partial \phi)^2 + e^{s_{\Omega} \phi} \frac{4n(n+1)}{(2-nN)^2}  \nonumber \\
	&&  -\sum_I  \frac{4(D-2)}{(2-nN)^2 \Delta_I} e^{s_{F_I}  \phi} \bigg] \,,
\end{eqnarray}
with the coefficients
\begin{eqnarray}
 s_1 &=& \left(\frac{D}{2}-1\right)\omega_0 + \left(\frac{d_0}{2}-1\right)\omega_1 \,, \\
  s_2 &=& \left(\frac{D}{2}-1\right)\omega_0 + \left(\frac{d_0}{2}-1\right)\omega_1 + \left(\frac{d_1}{2}-1\right)\omega_2 \,, \\
 s_{\Omega} &=& \left(\frac{D}{2}-1\right)\omega_0 + \frac{d_0}{2}\omega_1 + \frac{d_1}{2}\omega_2 + \frac{d_2}{2}\omega_{\text{E}} \,, \\
 s_{\phi} &=&  \frac{1}{2} - \frac{(D-2)(D-1)}{4}\omega_0^2 - \lambda_1 - \lambda_2 - \lambda_{\text{E}} \,, \\
 s_{F_1} &=& a_1 + \left(\frac{D}{2}-(q_1+2) \right)\omega_0 + \left(\frac{d_0}{2}-(q_1+2) \right) \omega_1  \\
 &&  +\left(\frac{d_1}{2}-(\bar{q}+2) \right) \omega_2 + \left(\frac{d_2}{2}-(\bar{q}+2) \right) \omega_{\text{E}} + 2b \,, \\
 s_{F_2} &=& a_2 + \left(\frac{D}{2}-(q_2+2) \right)\omega_0 + \sum_{I=1}^2 \left(\frac{d_{I-1}}{2}-(q_2+2) \right) \omega_I  \\
 && +\left(\frac{d_2}{2}-(\bar{q}+2) \right) \omega_{\text{E}} + 2b \,, \\
 \lambda_1 &=& - \frac{(d_0-2)(d_0-1)}{4}\omega_1^2 + (d_0+1)\omega_1 s_1  \,, \\
 \lambda_2 &=& - \frac{(d_1-2)(d_1-1)}{4}\omega_2^2 + (d_1+1)\omega_2 s_2  \,, \\
 \lambda_{\text{E}} &=& - \frac{(d_2-2)(d_2-1)}{4}\omega_{\text{E}}^2 \,, \\
 b &=& - \left(\bar{q} + 1 + \frac{2n}{2 - nN} \right) \frac{1}{\alpha} \,.
\end{eqnarray}
We can now perform the Kaluza-Klein reduction to go to a ($\bar{q}+1$)-dimensional theory. The reduction is done along the $x^{+}$ direction. The Kaluza-Klein decomposition ansatz in the Einstein frame is
\begin{equation}
	d\hat s^2_{\text{E}}= e^{\frac{1}{(\bar q-1)\bar q} \varphi} ds^2_{\text{E}} + e^{ -\frac{1}{\bar q}\varphi }( dx^{+} + A_{\mu} dx^{\mu} )^2 \,.
\end{equation}
Note that one must require that $\bar{q} \geq 2$. The action is
\begin{eqnarray}
	S &=& \int d^{\bar{q}+1}x \sqrt{- g } \bigg[ R - \frac{1}{4(\bar{q}-1)\bar{q}} (\partial \varphi)^2 + s_{\phi} (\partial \phi)^2 + e^{s_{\Omega} \phi + \frac{1}{(\bar{q}-1)\bar{q}} \varphi} \frac{4n(n+1)}{(2-nN)^2}  \nonumber \\
	&& - \sum_I \frac{4(D-2)}{(2-nN)^2 \Delta_I} e^{s_{F_I}\phi + \frac{1}{(\bar{q}-1)\bar{q}} \varphi} - \frac{1}{4} e^{-\frac{1}{\bar{q}-1} \varphi}F^2 \bigg] \,,
\end{eqnarray}
The Kaluza-Klein dilaton is
\begin{equation}
	\varphi = \bar{q}
	\left(\Omega_2 + \Omega_{\text{E}} + \frac{2n}{2-nN} + 2 \right) \log \rho \,,
\end{equation}
and the final space-time is $\bar{q} + 1$ dimensional with metric,
\begin{equation} \label{eqn:intbranefinal}
	ds^2_{\bar{q}+1} = \rho^{\kappa} \left( - \rho^{\frac{2n}{2-nN}} (dx^{-})^2 + d\rho^2 + d\vec{x}^2 \right) \,,
\end{equation}
where
\begin{equation} \label{eq:kappaBrane}
	\kappa = \frac{\bar{q}}{1-\bar{q}} \left( \Omega_2 + \Omega_{\text{E}} +2 \right) + \frac{1}{1-\bar{q}} \left(  \frac{2n}{2-nN} \right) \,,
\end{equation}
and finally the KK vector is
\begin{equation}
 \mathcal{A}  = \rho^{\frac{2n}{2-nN}} dx^{-} \,.
\end{equation}
Using the expressions for $\Omega_2$ and $\Omega_{\text{E}}$ given by \eqref{eqn:OmegaI} and \eqref{eqn:OmegaE} we find that $\kappa$ in \eqref{eq:kappaBrane} simplifies, and we arrive at the metric given in \eqref{eq:finalintersectingbranes}.

\addcontentsline{toc}{section}{References}
\bibliographystyle{newutphys}
\bibliography{LifshitzHR}

\end{document}